\documentclass{SciPost}

\hypersetup{
    colorlinks,
    linkcolor={red!50!black},
    citecolor={blue!50!black},
    urlcolor={blue!80!black}
}

\usepackage[bitstream-charter]{mathdesign}
\DeclareSymbolFont{usualmathcal}{OMS}{cmsy}{m}{n}
\DeclareSymbolFontAlphabet{\mathcal}{usualmathcal}

\fancypagestyle{SPstyle}{
\fancyhf{}
\lhead{\colorbox{scipostblue}{\bf \color{white} ~SciPost Physics }}
\rhead{{\bf \color{scipostdeepblue} ~Submission }}

\fancyfoot[C]{\textbf{\thepage}}
}

\usepackage{changes}
\usepackage[ruled,vlined]{algorithm2e}
\usepackage{hyperref} 
\usepackage{graphicx}       
\usepackage{dcolumn}        
\usepackage{bm}             
\usepackage{braket}         
\usepackage{color}
\usepackage{subcaption}
\usepackage{multirow}
\usepackage{appendix}
\usepackage{comment}
\usepackage[utf8]{inputenc}
\usepackage[capitalise]{cleveref}
\usepackage{enumitem}
\usepackage[inkscapelatex=false]{svg}
\hypersetup{
    colorlinks=true,
    linkcolor=blue,
    filecolor=magenta,      
    urlcolor=cyan,
    pdfpagemode=FullScreen,
}

\definecolor{ROYALBLUE}{rgb}{0.25, 0.41, 0.88}
\definecolor{ROYALPURPLE}{rgb}{0.47, 0.32, 0.66}
\definecolor{eggplant}{RGB}{142,18,156}
\definecolor{darkgreen}{RGB}{1,113,0}

\newcommand{\ms}[1]{{\color{eggplant}{#1}}}
\newcommand{\msc}[1]{{\color{eggplant}{[MS: #1]}}}
\newcommand{\zm}[1]{{\color{orange}{#1}}}
\newcommand{\zmc}[1]{{\color{darkgreen}{[ZM: #1]}}}

\newcommand{\jlc}[1]{{\color{purple}{[JL: #1]}}}

\newcommand{\refplaceholder}[1]{{\color{orange}{[ref]}}}

\begin{document}

\pagestyle{SPstyle}

\begin{center}{\Large \textbf{\color{scipostdeepblue}{
Recursive Sketched Interpolation: \\Efficient Hadamard Products of Tensor Trains\\
}}}\end{center}

\begin{center}\textbf{
Zhaonan Meng\textsuperscript{1$\star$},
Yuehaw Khoo\textsuperscript{2}, 
Jiajia Li\textsuperscript{1} and 
E. Miles Stoudenmire\textsuperscript{3$\dagger$}
}\end{center}

\begin{center}
{\bf 1} Department of Computer Science, North Carolina State University, Raleigh, NC 27695, USA
\\
{\bf 2} CCAM and Department of Statistics, University of Chicago, Chicago, IL 60637, USA.
\\
{\bf 3} Center for Computational Quantum Physics, Flatiron Institute, New York, NY 10010, USA
\\[\baselineskip]
$\star$ \href{mailto:email1}{\small zmeng5@ncsu.edu}\,,\quad
$\dagger$ \href{mailto:email2}{\small mstoudenmire@flatironinstitute.org}
\end{center}

\section*{\color{scipostdeepblue}{Abstract}}
\textbf{\boldmath{%
The Hadamard product of two tensors in the tensor-train (TT) format is a fundamental operation across various applications, such as TT-based function multiplication for nonlinear differential equations or convolutions. However, conventional methods for computing this product typically scale as at least $\mathcal{O}(\chi^4)$ with respect to the TT bond dimension (TT-rank) $\chi$, creating a severe computational bottleneck in practice. By combining randomized tensor-train sketching with slice selection via interpolative decomposition, we introduce Recursive Sketched Interpolation (RSI), a ``scale product'' algorithm that computes the Hadamard product of TTs at a computational cost of $\mathcal{O}(\chi^3)$. Benchmarks across various TT scenarios demonstrate that RSI offers superior scalability compared to traditional methods while maintaining comparable accuracy. We generalize RSI to compute more complex operations, including Hadamard products of multiple TTs and other element-wise nonlinear mappings, without increasing the complexity beyond $\mathcal{O}(\chi^3)$.
}}

\vspace{\baselineskip}

\noindent\textcolor{white!90!black}{%
\fbox{\parbox{0.975\linewidth}{%
\textcolor{white!40!black}{\begin{tabular}{lr}%
  \begin{minipage}{0.6\textwidth}%
    {\small Copyright attribution to authors. \newline
    This work is a submission to SciPost Physics. \newline
    License information to appear upon publication. \newline
    Publication information to appear upon publication.}
  \end{minipage} & \begin{minipage}{0.4\textwidth}
    {\small Received Date \newline Accepted Date \newline Published Date}%
  \end{minipage}
\end{tabular}}
}}
}


\vspace{10pt}
\noindent\rule{\textwidth}{1pt}
\tableofcontents
\noindent\rule{\textwidth}{1pt}
\vspace{10pt}


\section{Introduction}

Many problems in the natural sciences, ranging from fluid dynamics to quantum chemistry, rely heavily on numerical simulations. A significant portion of the computational effort in these simulations is devoted to resolving multiscale continuum features, tracking the positions of many particles, or capturing correlations among multiple variables.
These seemingly distinct challenges can be brought into a unified framework by representing the state of the system as a tensor. In this representation, groups of tensor indices correspond to specific particles or variables, or encode length scales to capture continuum structures.

However, representing a high-dimensional function directly by a tensor is impractical due to the \emph{curse of dimensionality}~\cite{cichocki2016tensor, oseledets2009breaking}: The number of tensor components grows exponentially with the number of indices (its order), leading to prohibitive computational and memory costs. This obstacle can be overcome for many problems through tensor networks~\cite{orus2014practical, cichocki2016tensor, orus2019tensor}, a class of efficient representations that approximate high-order tensors as contractions of simpler, lower-order tensors. By maintaining low ranks, also known as bond dimensions---the sizes of the internal indices that connect the factors in a tensor network—one can achieve exponential reductions in both memory requirements and computational complexity. One of the most fundamental tensor networks is the matrix product state (MPS)~\cite{fannes1992finitely, klumper1992groundstate, ostlund1995thermodynamic, vidal2003efficient}, equivalently referred to as the tensor train (TT)~\cite{Oseledets2011Tensor-TrainDecomposition}, in which a tensor is represented as the contraction of a chain of third-order tensors called TT-cores.

Numerous efficient algorithms based on tensor trains have been developed, such as the density matrix renormalization group (DMRG) algorithm~\cite{white1992density,schollwock2011density}, which computes the eigenvectors of operators represented as tensor networks. While most algorithms focus on efficient linear operations of TT for solving linear equations or eigenvalue problems, nonlinear TT operations such as the Hadamard product, which performs element-wise multiplication of tensors in TT format, are also essential. The Hadamard product is critical for TT-based numerical methods, such as solving nonlinear differential equations like the Navier–Stokes equations in fluid mechanics~\cite{gourianov2022quantum, peddinti2024quantum} or the Gross–Pitaevskii equation for quantum systems~\cite{boucomas2025quanticstensortrainsolving, SciPostPhys.21.2.035}. Other tasks, ranging from computing interaction integrals in quantum chemistry \cite{Jolly_2025} to using tensor networks for optimization, also benefit from fast product operations. However, performing the Hadamard product of TTs is nontrivial, as individual tensor entries are not directly accessible in the TT format, and conventional methods to compute it exhibit unfavorable scaling with respect to the TT bond dimension. 

Given two TTs with bond dimension $\chi$, the direct method for computing their Hadamard product is forming the Kronecker product of their respective TT-cores~\cite{Oseledets2011Tensor-TrainDecomposition}. 
This requires at least $\mathcal{O}(\chi^4)$ operations and yields an output TT with bond dimension $\chi^2$. 
The resulting bond dimension is typically redundant, necessitating a rank-compression step such as TT‑rounding~\cite{Oseledets2011Tensor-TrainDecomposition, Sun_2024}. Due to the $\chi^2$ rank, the rounding step has even higher complexity, dominating the total computational overhead~\cite{Sun_2024}.
Other methods for computing the product also exist, including classical MPO-MPS contraction algorithms such as the variational and zip-up methods~\cite{schollwock2011density,Verstraete2004MPO,Stoudenmire_2010}, cross-interpolation-based approaches~\cite{Oseledets2010TT-crossArrays,NunezFernandez2025LearningLibraries}, and more recent techniques~\cite{Sun_2024, michailidis2025element}. Nevertheless, many of these still scale as at least $\mathcal{O}(\chi^4)$ or generate impractically large intermediate bond dimensions. 


\begin{figure}[htp]
    \centering
    \includegraphics[width=0.7\columnwidth]{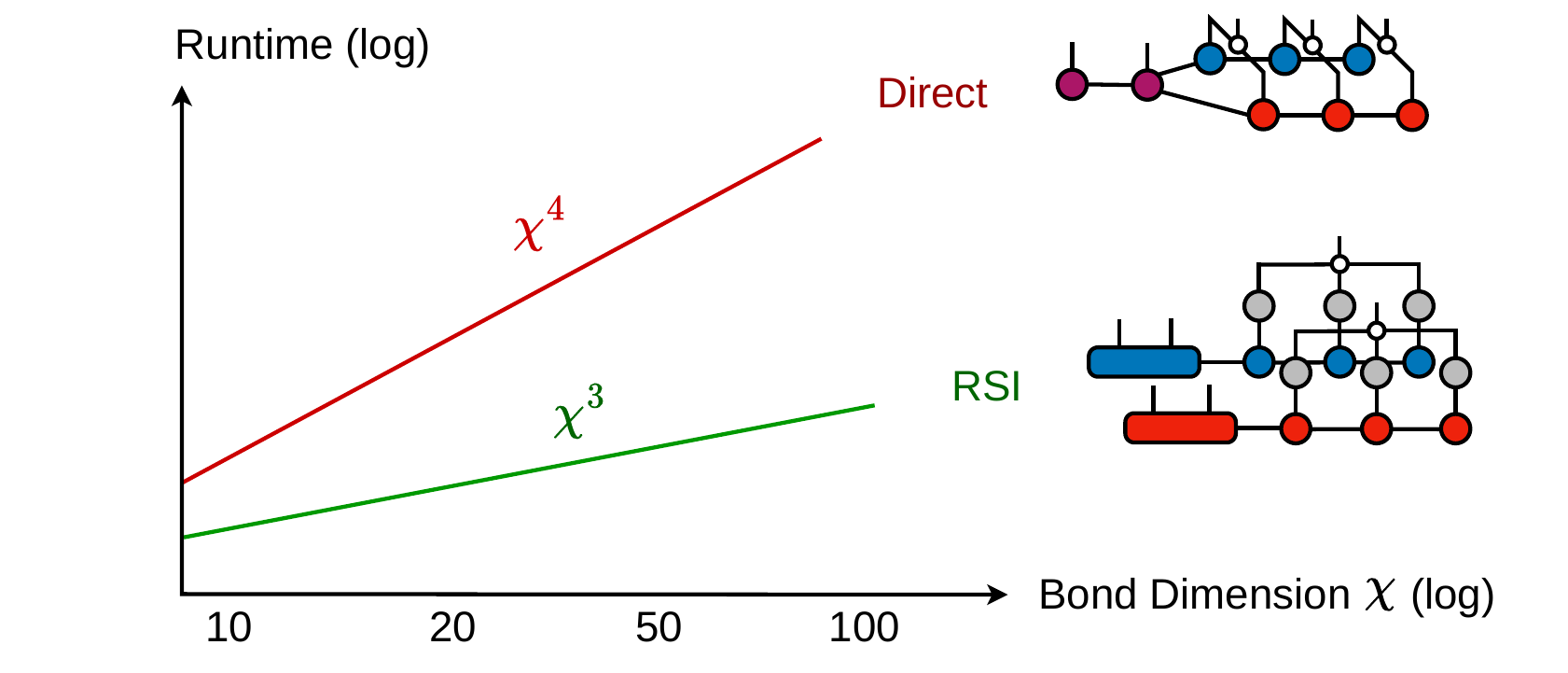}
    \caption{\raggedright The recursive sketched interpolation method offers a $\chi^3$ runtime complexity versus the direct method which scales at least as $\chi^4$. The tensor diagrams illustrate snapshots of the respective computational procedures.}
    \label{fig: ID diagram}
\end{figure}

Motivated by the scalability limits of existing TT multiplication methods, we propose a novel ``scale product'' algorithm, \emph{Recursive Sketched Interpolation} (RSI). Maintaining a compressed bond dimension $\chi$ throughout the entire computation, RSI achieves a reduced computational complexity of $\mathcal{O}(\chi^3)$.
RSI employs two key techniques: tensor-train sketching for dimensionality reduction and range extraction, and interpolative decomposition to obtain low-rank factorizations while preserving access to tensor entries in the defining basis.
Our experimental results across various TT scenarios demonstrate that RSI achieves superior runtime scalability with respect to bond dimension compared to conventional methods, while maintaining comparable approximation accuracy.
We further show the extensibility of RSI to more complex element‑wise operations, including the Hadamard product of more than two TTs and other nonlinear mappings, without increasing the computational complexity beyond $\mathcal{O}(\chi^3)$.

The paper is organized as follows. \Cref{sec: Background} covers the background necessary for this work. \Cref{sec: RSI Algorithm} introduces the RSI algorithmic framework, discussing its implementation details and computational complexity. In \Cref{sec: Numerical Experiments}, we evaluate the accuracy and runtime performance of RSI and compare it with conventional baseline methods through various numerical experiments. Following the experiments, \Cref{sec: applications} briefly demonstrates applications of RSI. And \Cref{sec: nonlinear map of rsi} discusses the extensibility of RSI by generalizing it to compute other element-wise nonlinear mappings of tensor trains. Finally, \Cref{sec: conclusion} concludes the paper with a summary of RSI and future research directions.

\section{Background}
\label{sec: Background}

This section reviews essential background on tensor trains, interpolative decomposition, randomized sketching, and related work on Hadamard products of TTs.

\subsection{Tensor Networks and Diagram Notation}

A tensor network provides an efficient and interpretable representation of high-order tensors by expressing them as contractions of multiple lower-order tensors, where \emph{order} denotes the number of indices (i.e.\ dimensionality of the tensor). Tensor networks have become a central tool for efficiently manipulating tensors that would otherwise be prohibitively large. Various formats have been widely studied, including the matrix product state (MPS)~\cite{fannes1992finitely, klumper1992groundstate, ostlund1995thermodynamic, vidal2003efficient}, the projected entangled pair state (PEPS)~\cite{verstraete2004renormalization, verstraete2004valence}, and tree tensor networks~\cite{shi2006classical}. 

\begin{figure}[h]
    \centering
    \begin{subfigure}{0.4\textwidth}
        \centering
        \includegraphics[width=0.5\textwidth]{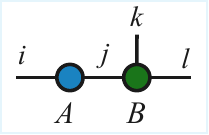}
        \caption{The contraction $\sum_jA_{ij}B_{jkl}$.}
        \label{fig: AB contraction diagram}
    \end{subfigure}
    \begin{subfigure}{0.5\textwidth}
        \centering
        \includegraphics[width=1.0\textwidth]{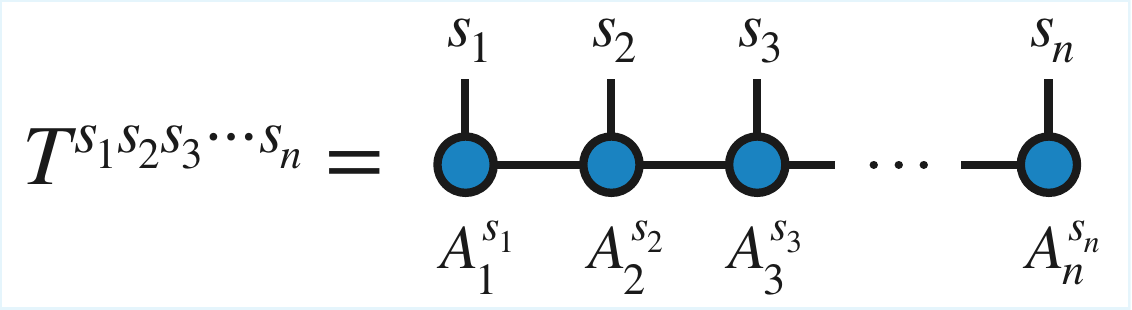}
        \caption{Tensor train of the tensor $T^{s_1s_2,\ldots,s_n}$.}
        \label{fig: T MPS diagram}
    \end{subfigure}

    \vspace{1em}
    
    \begin{subfigure}{0.5\textwidth}
        \centering
        \includegraphics[width=0.8\textwidth]{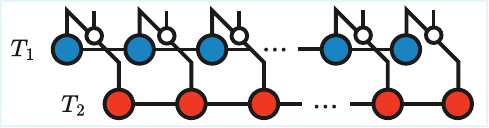}
        \caption{Hadamard product of two tensors in TT format computed via the core-wise Kronecker product. The white circles connecting the physical indices are copy tensors, $\delta_{ijk}$.}
    \label{fig: directmethod_diagram}
    \end{subfigure}
    
    \caption{Tensor diagram notation.} 
    \label{fig:main tensor diagram}
\end{figure}

The matrix product state, also known as the tensor train (TT), is one of the most common tensor network formats. A wide range of controlled algorithms have been developed for tensor trains, such as computing sums or inner products or solving linear equations when the solution can be represented in TT form. 
An order-$n$ tensor $T^{s_1 s_2 \ldots s_n}$ can be represented as a tensor train as follows:
\begin{equation}
\label{eq: MPS format}
T^{s_1 s_2 \cdots s_n}= \sum_{\alpha_1=1}^{\chi_1} \sum_{\alpha_2=1}^{\chi_2}\cdots \sum_{\alpha_{n-1}=1}^{\chi_{n-1}} A^{s_1}_{1,\alpha_1} A^{s_2}_{\alpha_1,\alpha_2} \cdots A^{s_j}_{\alpha_{j-1},\alpha_j} \cdots A^{s_n}_{\alpha_{n-1},1},
\end{equation}
where each tensor $A$, referred to as a \emph{TT-core}, is contracted with its neighbors over the shared bond indices $\alpha_j$ of dimension $\chi_j$. Figure~\ref{fig:main tensor diagram}(b) shows a TT decomposition using diagrammatic notation. The values $\{\chi_j\}_{j=1}^{n-1}$ are the \emph{bond dimensions} (or \emph{TT-ranks}) of the TT, and the indices $\{s_j\}_{j=1}^n$ are the \emph{physical indices}, whose sizes constitute the \emph{physical dimensions} $\{d_j\}_{j=1}^n$. Note that all $A^{s_j}$ can all be different from one another---it is common to sometimes use the same letter for each core when the cores are understood to be identified by the indices they carry. 

If the bond dimension is large enough, a tensor train can exactly represent an arbitrary tensor. For an order-$n$ tensor with uniform physical dimensions $\{d_j\}=d$, an exact TT representation always exists with a maximum bond dimension of $\chi = d^{\lfloor n/2 \rfloor}$. In most practical applications, however, the TT is used as an approximation, with the bond dimension truncated below a fixed, moderate value or determined adaptively. The intermediate TT-cores are third-order tensors of size $\chi_{j-1} \!\times\! d \!\times\! \chi_j$, except for the two boundary factors $A^{s_1}$ and $A^{s_n}$, which are matrices of sizes $d \!\times\! \chi_1$ and $\chi_{n-1} \!\times\! d$, respectively. The TT compresses the size complexity of storing the tensor $T$ from $\mathcal{O}(d^n)$ to $\mathcal{O}(n d \chi^2)$, leading to a great reduction in the number of parameters in tensor processing.

Tensor networks can be represented using a graphical notation, known as tensor diagram notation~\cite{penrose1971applications, cvitanovic2008group}. The diagrammatic notation represents contractions between factors, such as the operation $\sum_j A_{ij} B_{jkl}$ or the tensor train from Eq.~(\ref{eq: MPS format}), as depicted in \Cref{fig:main tensor diagram}. Notably, the Hadamard product of  TTs can also be represented by a tensor diagram using a \emph{copy} tensor~\cite{ahle2024tensorcookbook, boucomas2025quanticstensortrainsolving}. The copy tensor $\delta_{ijk\ldots}$, with dimensions $i,j,k,\ldots$, denotes a diagonal tensor with entries equal to 1 if all indices are identical and 0 otherwise:
\begin{equation} 
\delta_{ijk\ldots} = 
\begin{cases} 
1 & \text{if } i = j = k = \cdots, \\
0 & \text{otherwise.} 
\end{cases} 
\end{equation}
Utilizing the diagrammatic notation of the third-order copy tensor $\delta_{ijk}$—depicted as a white circle $\circ$—we can express the TT Hadamard product in a form equivalent to the direct method using core-wise Kronecker products---see Fig.~\ref{fig:main tensor diagram}(c).

\subsection{Interpolative Decomposition}
\label{sec: interpolative decomposition}

Low-rank approximation methods, such as the singular value decomposition (SVD), have been extensively studied for both matrices and tensors. 
However, other low-rank factorizations exist besides SVD which trade slightly higher ranks for other benefits.
The interpolative decomposition (ID)~\cite{cheng2005compression, Liberty2007RandomizedMatrices,martinsson2020randomized} decomposes a matrix into two factor matrices that are constructed from a subset of its own rows or columns. Given an $m \!\times\! n$ matrix $M$, a rank-$k$ \emph{row ID} of $M$ yields the approximation $M \!\approx\! X R$, where
\begin{itemize}[leftmargin=*]
\item $R=M(\mathcal{I}, :)$ is a $k \!\times\! n$ matrix consisting of $k$ rows $\mathcal{I} \subset \{1, \dots, m\}$ of $M$. $R$ is termed the row \emph{skeleton} or \emph{pivots}, while the index set $\mathcal{I}$ denotes the row skeleton indices.
\item $X$ is an $m \!\times\! k$ \emph{interpolation} matrix containing a $k \!\times\! k$ identity submatrix. The remaining $m-k$ rows are assembled into a coefficient matrix that expresses the non-selected rows as linear combinations of the skeleton rows.
\end{itemize}

Similarly, an analogous form applies to \emph{column ID}, yielding the approximation $M \!\approx\! CZ$ with column skeleton $C = M(:, \mathcal{J})$, while \emph{double-sided ID} yields $M \!\approx\! X M_s Z$ with the core submatrix $M_s = M(\mathcal{I}, \mathcal{J})$ formed by the intersection of the row and column skeletons. Beyond the ID format, a related representation is the \emph{cross} decomposition~\cite{goreinov1997theory, goreinov1997pseudo, chiu2013sublinear}, where $M \!\approx\! CUR$ with column and row skeletons $C$ and $R$, and middle matrix $U = M_s^{-1}$ for a matrix of exact rank $k$. 
The cross format is related to the ID format by absorbing $U$ into the skeleton matrices, as illustrated in \Cref{fig: ID diagram}.
\begin{figure}[htp]
    \centering
    \includegraphics[width=0.85\columnwidth]{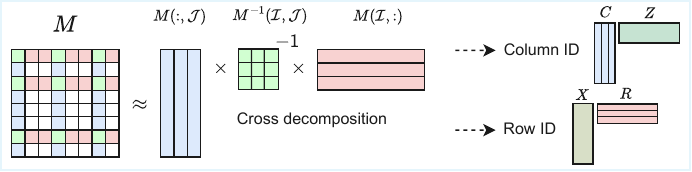}
    \caption{\raggedright Cross and interpolative decomposition of a matrix $M$. The cross format can reduce to a one-side ID form. Note that ID factorizations can be computed directly without passing through the cross form.}
    \label{fig: ID diagram}
\end{figure}

In the ID approximation, the entries in the skeleton rows $\mathcal{I}$ and columns $\mathcal{J}$ are reproduced exactly, whereas the rest are interpolated from the pivots. The quality of the factorization depends on how well the selected $\mathcal{I}$ and $\mathcal{J}$ span the range of the matrix. The selection of optimal pivots of a matrix is typically heuristic and often guided by the maximum volume principle, inspiring several methods, including rank-revealing decompositions~\cite{CHAN198767, PAN2000199, NunezFernandez2025LearningLibraries}, nuclear-score maximization~\cite{fornace2024columnrowsubsetselection}, among others.

Compared to SVD, which gives the best low-rank approximation in terms of spectral norm, ID trades orthogonality and some accuracy for the advantage of representing the full matrix by its own entries. 
This advantage motivates the generalization of ID from matrices to tensors, yielding the tensor cross interpolation algorithm (TCI, or TT-cross)~\cite{NunezFernandez2025LearningLibraries, Oseledets2010TT-crossArrays}. TCI constructs a tensor-train approximation of a tensor by iteratively applying ID to the tensor slices. By evaluating only partial tensor entries during the heuristic pivot search in each dimension, TCI constructs a TT without explicitly accessing all tensor elements, enabling the efficient approximation of very high-dimensional tensors.

\subsection{Randomized Sketching}
\label{sec: background TT sketching}


Randomized linear embedding, a cornerstone of randomized numerical linear algebra, utilizes randomness to construct compressed surrogates for large linear operators~\cite{woodruff2014sketching, martinsson2020randomized}. The core principle is to apply a random linear map, often referred to as a \emph{sketch}, to project high-dimensional data onto a lower-dimensional subspace while preserving essential geometric properties with high probability~\cite{johnson1984extensions}. 
By performing such compression, sketching allows large-scale problems to be solved via their lower-dimensional counterparts with provable error guarantees.
Sketching has become a widely used approach for the efficient computation of large-scale matrix and tensor operations across diverse scientific domains, notably for randomized low-rank approximations like the randomized SVD~\cite{Halko2011RandomizedMatrixDecomp}.

Inspired by the success of randomized methods in matrix algebra, sketching techniques have also been developed for tensor networks such as tensor trains. 
For instance, Hur et al.~\cite{hur2023generativemodelingtensortrain} demonstrate that sketching is equally powerful for generative modeling, enabling the recovery of tensor trains from independent samples by solving sketched linear systems, effectively bypassing the curse of dimensionality in density estimation tasks. In the context of tensor contraction, Camaño et al.~\cite{camano2025successiverandomizedcompressionrandomized} apply similar principles to the matrix product operator-matrix product state (MPO-MPS) product; their \emph{successive randomized compression} algorithm performs single-pass adaptive compression, avoiding the high computational cost of variational sweeps. Daas et al.~\cite{daas2025adaptiverandomizedtensortrain} utilize sketches to facilitate adaptive TT-rounding, establishing a rigorous error estimator that allows for huge speedups compared to deterministic approaches.

\subsection{Previous Methods for Hadamard Product of TTs}
\label{sec: other TT product methods}

The conventional approach for the Hadamard product of two TT tensors is to form the Kronecker product of their corresponding cores \cite{Oseledets2011Tensor-TrainDecomposition}, which we refer to as the ``direct'' method. While straightforward and exact, computing the Kronecker product scales as the fourth power of the input bond dimension and yields a TT whose bond dimensions equal the product of the inputs, which is usually highly redundant. Although existing literature explores reducing the cost of TT-rounding for rank truncation~\cite{Sun_2024, daas2025adaptiverandomizedtensortrain, al2023randomized}, the subsequent rounding process remains time-consuming, and the direct method still incurs a minimum complexity of $\mathcal{O}(\chi^4)$ in the bond dimension $\chi$. 




Besides the direct construction and truncation, classical tensor-network algorithms can also be employed. In particular, the variational and zip-up methods~\cite{schollwock2011density,Verstraete2004MPO,Stoudenmire_2010}, which are commonly used for MPO--MPS contractions, can be adapted to compute Hadamard products by viewing one TT as a diagonal MPO~\cite{peddinti2024quantum, schollwock2011density}. The zip-up method constructs the output TT sequentially through local contractions and truncated SVD. The variational method formulates the contraction as an optimization problem and iteratively sweeps over the TT-cores to minimize the approximation error. Both methods yield a product with compressed bond dimensions when applying the MPO-style TT to the other TT. Nevertheless, the dominant contractions in these approaches involve four virtual bond indices associated with the two input TTs, still leading to a $\mathcal{O}(\chi^4)$ complexity.

Another known approach is the tensor cross interpolation (TCI) method. TCI evaluates slices of the product tensor from input TTs and employs cross approximation through back-and-forth iterations to interpolate the resulting TT. While TCI can approximate any element-wise nonlinear mapping of input TTs into a target TT, its efficiency remains constrained by its scalability. 
With a target output bond dimension of $\chi$, every cross-interpolation step queries $\mathcal{O}(\chi^2)$ entries of the output tensor. Since we assume the inputs are only known in TT form, if each query is evaluated independently by contracting the input TTs, it incurs an additional $\mathcal{O}(\chi^2)$ cost, and the overall cost scales as $\mathcal{O}(\chi^4)$. In the latest advances, this complexity can be reduced by batch evaluation, where the input-TT environments over the left and right pivots are built recursively and shared across all queries of a slice~\cite{NunezFernandez2025LearningLibraries,tttoolbox}.



Beyond conventional methods, several recent approaches have also been proposed. The recent work by Michailidis et al.~\cite{michailidis2025element} presented an early, effective attempt based on multiplication between a pair of two TT-cores followed by core swapping to iteratively process the next core pair. Concurrently, Sun et al.~\cite{Sun_2024} introduced the Hadamard‑avoiding TT recompression (HaTT) algorithm, which aims to avoid the high cost of TT‑rounding in the product computation. 
While these methods have shown promising results in various tests, we notice that they may face challenges such as large intermediate bond dimensions or a conditional complexity that can revert to $\mathcal{O}(\chi^4)$ when the target output bond dimension equals the input $\chi$.

\section{Recursive Sketched Interpolation Algorithm}
\label{sec: RSI Algorithm}

In this section, we present and analyze Recursive Sketched Interpolation (RSI), a novel algorithm for computing Hadamard products of tensors in the tensor-train format. 

\subsection{Hadamard Product of TTs via RSI}
\label{sec: RSI - Product of two tensor trains}
\begin{figure}[htb]
    \centerline{\includegraphics[width=1\textwidth]{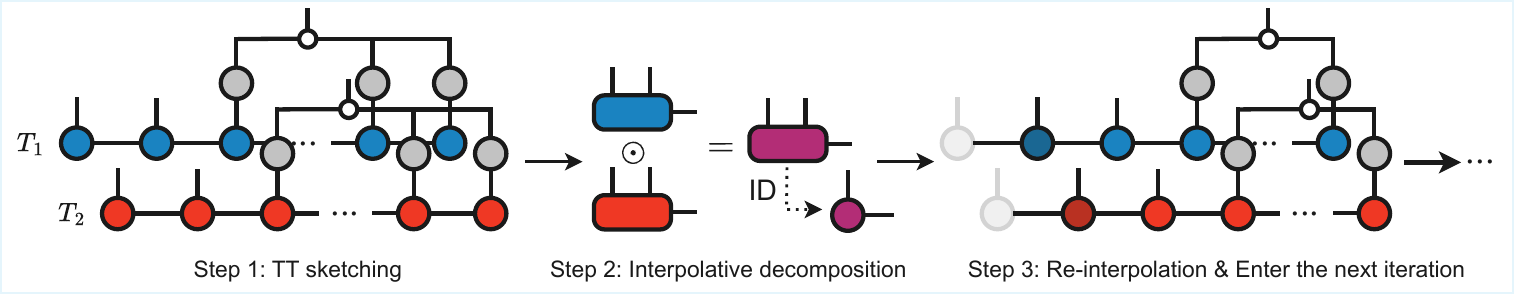}}
    \caption{Workflow overview of the first iteration of RSI.}
    \label{fig: workflow rsi 1st iter}
\end{figure}

Consider two order-$n$ tensors $T_1$ and $T_2$  of identical external dimensions $\{d_j\}_{j=1}^n$, expressed in the TT format as 
\begin{align}
T_1^{s_1s_2\cdots s_n}=\sum_{\alpha_1=1}^{\chi_1^A} \sum_{\alpha_2=1}^{\chi_2^A}\!\cdots\! \sum_{\alpha_{n-1}=1}^{\chi_{n-1}^A} A^{s_1}_{1,\alpha_1} A^{s_2}_{\alpha_1,\alpha_2} \!\cdots\! A^{s_n}_{\alpha_{n-1},1},\text{ }
T_2^{s_1s_2\cdots s_n}=\sum_{\alpha_1=1}^{\chi_1^B} \sum_{\alpha_2=1}^{\chi_2^B}\!\cdots\! \sum_{\alpha_{n-1}=1}^{\chi_{n-1}^B} B^{s_1}_{1,\alpha_1} B^{s_2}_{\alpha_1,\alpha_2} \!\cdots\! B^{s_n}_{\alpha_{n-1},1}.
\end{align}
We make no assumptions about the initial ``\emph{gauge freedom}'' of the tensor train---that is, the freedom to reparameterize the TT-cores without altering the values of the tensor. The goal of the algorithm is to perform the Hadamard product of $T_1$ and $T_2$ from their TT representations. This operation defines a tensor $G$ as the element-wise product of $T_1$ and $T_2$, denoted by $G = T_1\odot T_2$, such that
\begin{align}
G^{s_1 s_2 s_3 \cdots s_n} \equiv T^{s_1 s_2 s_3 \cdots s_n}_1 T^{s_1 s_2 s_3 \cdots s_n}_2  
\end{align}
for all index values ${s_1, s_2, \ldots, s_n}$. RSI constructs a TT approximation of $G$ by iteratively sketching the input TTs and applying interpolative decomposition to the Hadamard product of the sketched tensors, generating one TT‑core per iteration from index $1$ to $n$. \Cref{fig: workflow rsi 1st iter} illustrates the first iteration of RSI computing the initial TT-core of $G$, with each step explained in detail in the following subsections. The remaining TT-cores are then obtained by repeating the procedure recursively.

\subsubsection{Step 1---Tensor-Train Sketching}
\label{sec: step1 skecth TT}
The RSI iteration starts with a dimensionality reduction step by randomized TT sketching. Both TTs undergo the equivalent sketching, and for convenience we describe the procedure only for $\{A^{s_j}\}_{j=1}^n$ here. For the TT $\{A^{s_j}\}$, we leave the first two external indices open and contract the remaining indices with a random tensor $S^{k}_{s_3 \cdots s_n}$, resulting in a third-order tensor $\tilde{T}_1$. The tensor $S$ is called the \emph{sketch}, and the size of its index $k$ is referred to as the \emph{sketch dimension}, which we also denote simply by $k$.
\begin{center}
\includegraphics[width=0.52\columnwidth]{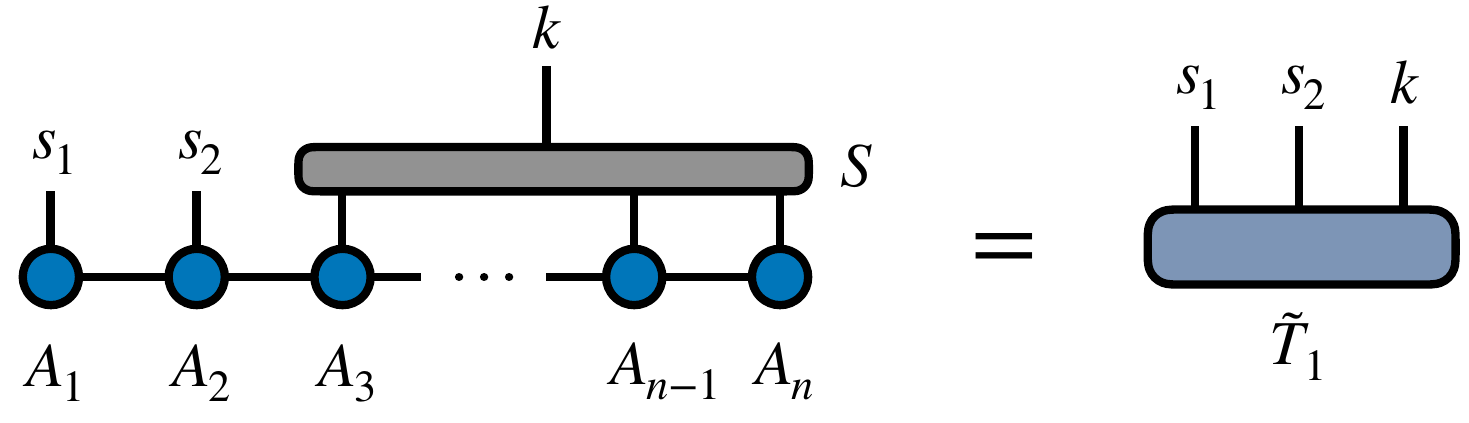}
\end{center}

Note that the sketch $S$ (the gray tensor in the above diagram) 
should not be formed explicitly, as applying the full $S$ would be inefficient for both storage and computation and would reintroduce the curse of dimensionality. 
A suitable sketch should be chosen to be: (i) any tensor network, or a collection thereof, that can be applied to the TT efficiently; (ii) sufficiently generic so that $\tilde{T}$ spans the dominant subspace of $T$, when ``unfolding'' or ``matricizing'' $T$ into a matrix with $(s_1,s_2)$ as the row index and $(s_3, \ldots, s_n)$ as the column index. In what follows, we employ a \emph{random one-cluster basis} sketch~\cite{peng2023generativemodelinghierarchicaltensor, camano2025successiverandomizedcompressionrandomized}, which has been shown to offer a good balance between generality and efficiency. In physics terminology, this sketch could be described as computing partial overlaps with $k$ random product states supported on the indices $(s_3, \ldots, s_n)$.

For each index to be sketched, we generate independent random matrices $\Omega_{s_3}^k, \ldots, \Omega_{s_n}^k$ with entries sampled from the standard normal distribution. By contracting each $\Omega$ with its corresponding core tensor, we obtain the sketched TT-cores $\{\Omega^k_{s_j} A^{s_j}\}_{j=3}^n$ with an external dimension of $k$. 
For each value $s$ of the external index (with $s \!=\! 1, \!\ldots\!, k$), we fix the external index of every sketched core to $s$ and contract them to obtain a vector. This process yields $k$ vectors, assembled into a $\chi^A_2 \!\times\! k$ matrix sketching external indices $s_3,\ldots,s_n$ of the TT. The entire sketching operation can be depicted via the copy tensor $\delta_{s_3s_4\ldots s_n}$, as illustrated by the white circle $\circ_{s_3s_4\ldots s_n}$ in the diagram below:
\begin{center}
\includegraphics[width=0.9\columnwidth]{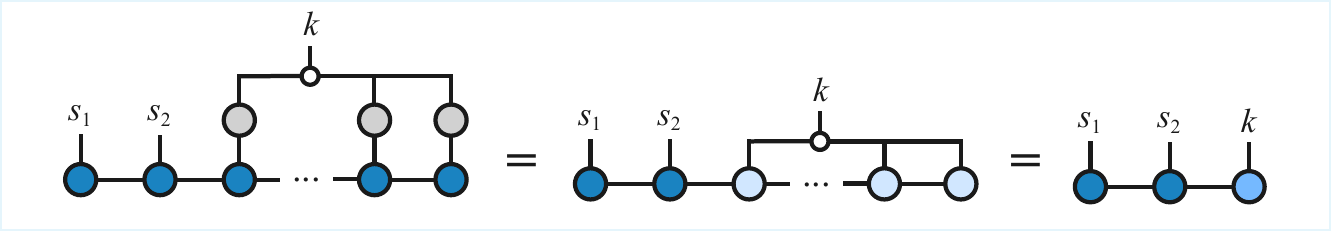}
\end{center}
In essence, the resulting $\chi^A_2 \!\times\! k$ \emph{sketch matrix} (light blue, at the far right) corresponds to the Khatri–Rao product (KRP) of all sketched TT-cores $\{\Omega^k_{s_j} A^{s_j}\}$~\cite{Kolda2009TensorDecompApp,camano2025successiverandomizedcompressionrandomized}. This sketching approach is very efficient because it avoids explicit construction of the full sketch and, moreover, enables the same sketch to be \emph{reused} in subsequent RSI iterations, as detailed in the recursion step in \Cref{sec: recursion of the algorithm}.

The resulting matrix sketches of $\{A^{s_j}\}$ and $\{B^{s_j}\}$ are then contracted with the first two index-open TT-cores, producing the sketched tensors $\tilde{T}_1^{s_1 s_2 k}$ and $\tilde{T}_2^{s_1 s_2 k}$.
\begin{center}
\includegraphics[width=0.9\columnwidth]{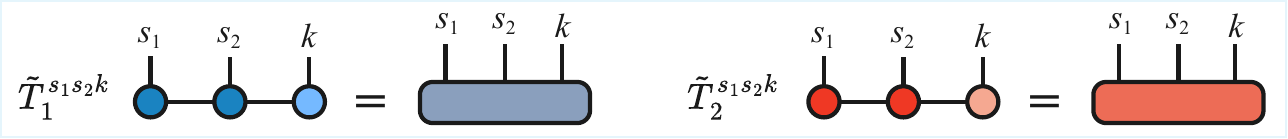}
\end{center}
Containing the compressed information from the $s_3$–$s_n$ TT-cores, $\tilde{T}_1^{s_1 s_2 k}$ and $\tilde{T}_2^{s_1 s_2 k}$ can be regarded as sketched approximations of the original tensors $T_1$ and $T_2$ at sites $s_1$ and $s_2$. With the $k$-space of $\tilde{T}_1$ and $\tilde{T}_2$ approximating the $(s_3,\dots,s_n)$-space of $T_1$ and $T_2$, we proceed to the next interpolative decomposition step.

\subsubsection{Step 2---Interpolative Decomposition of the Sketched Hadamard Product}

In the second step, RSI first computes the Hadamard product of the sketched tensors:
\begin{align}
\label{eq: sketch G=T1T2}
\tilde{G}^{s_1 s_2 k} = \tilde{T}_1^{s_1 s_2 k} \tilde{T}_2^{s_1 s_2 k},
\end{align}
\begin{center}
\includegraphics[width=0.51\columnwidth]{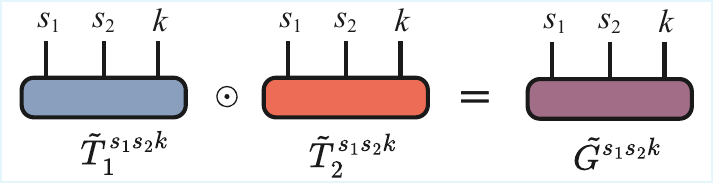}
\end{center}
which can be viewed as the sketched version of the output $G=T_1 \odot T_2$. The algorithm proceeds by decomposing $\tilde{G}$ with a row-based interpolative decomposition. This amounts to treating the first index $s_1$ as a ``row'' index and remaining indices $(s_2,k)$ as a grouped ``column'' index for the purpose of matricizing the tensor to compute the decomposition. The rank-$\chi$ row ID of the matricized $\tilde{G}^{s_1s_2k}$ yields
\begin{align}
\label{eq: row ID of G}
\tilde{G}^{s_1 [s_2 k]} \approx X^{s_1 \alpha_1}\, R^{\alpha_1 [s_2 k]},
\end{align}
\begin{center}
\includegraphics[width=0.47\columnwidth]{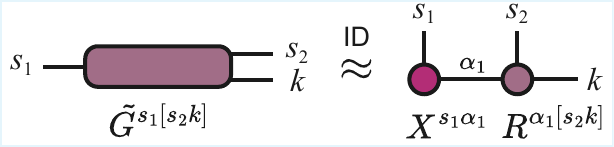}
\end{center}
where $\alpha_1$ is the bond index of dimension $\chi$. The matrix $R^{\alpha_1[s_2k]}$ is the row skeleton, formed by the $\chi$ rows of $\tilde{G}^{s_1 [s_2 k]}$ whose indices constitute the \emph{row pivot set} $\mathcal{I}_1$. The remaining rows of $\tilde{G}$ are approximated by linearly combining these pivots using the coefficients contained in the interpolation matrix $X^{s_1\alpha_1}$. The optimal pivots can be selected by the heuristic methods listed in \Cref{sec: interpolative decomposition}. In this work, we choose to use the partial rank-revealing LU method (prrLU), a well-established method used in tensor cross interpolation~\cite{NunezFernandez2025LearningLibraries,NunezFernandez2022LearningTrains}. In practice, we compute the ID until either the ID approximation error falls below the error tolerance $\epsilon_{\text{ID}}$ or the number of pivots reaches the user-defined maximum bond dimension $\chi_{\max}$.

Importantly, following the ID step, the matrix $R$ is discarded and not used in producing the final result, since its values are somewhat arbitrary as a result of the randomized sketching step. The interpolation matrix $X^{s_1\alpha_1}$ constitutes the first TT-core $X^{s_1}_{1,\alpha_1}$ of the TT approximation of $G$, with $\chi$ serving as the first bond dimension $\chi^G_1$. 
The underlying intuition behind using $X^{s_1}$ as the TT-core is that even though the sketched product $\tilde{G}^{s_1 s_2 k}$ does not match the true entries of $G$, it nonetheless contains sufficient structural information on how the true product $G = T_1 \odot T_2$ can be interpolated in the $s_1$ space. This enables $X^{s_1}$ to serve as an accurate interpolation of the $s_1$-space matricization of $G$, which then becomes the TT-core of $G$ at $s_1$.

\subsubsection{Step 3---Re-Interpolation of the Input TTs}
\label{sec: re-interpolation}



After computing the first core $X^{s_1}$ and its associated pivot set $\mathcal{I}_1$, which interpolates $G$ over index $s_1$, we prepare to compute the second core $X^{s_2}$. However, while $X^{s_2}$ is defined via nested interpolation from $X^{s_1}$ in the $s_1$ space, the residual input TT-cores ($A^{s_2}$–$A^{s_n}$, $B^{s_2}$–$B^{s_n}$) are not interpolated via $\mathcal{I}_1$ and thus interpolate through different slices of the $s_1$ space relative to $G$.
Therefore, before proceeding, the input TT-cores must be \emph{re-interpolated} to align their $s_1$ basis with $\mathcal{I}_1$.

The following diagram illustrates how to carry out the re-interpolation in practice. Using the TT-cores $\{A^{s_j}\}$ of $T_1$ as the example, we first contract the first two cores to form $A^{s_1}A^{s_2}$. Next, we select slices of this contracted tensor by fixing the $s_1$ index to the values in the pivot set $\mathcal{I}_1$. The resulting tensor of slices serves as the re-interpolated TT-core at $s_2$, where the left bond index $\alpha_1$ is re-interpolated to $\tilde{\alpha}_1$ with dimension $\chi^G_1$. Note that no interpolative decomposition procedure is necessary at this step and only slicing is performed. An analogous procedure is applied to re-interpolate the TT $\{B^{s_j}\}$ for $T_2$.

\begin{center}
\includegraphics[width=0.8\columnwidth]{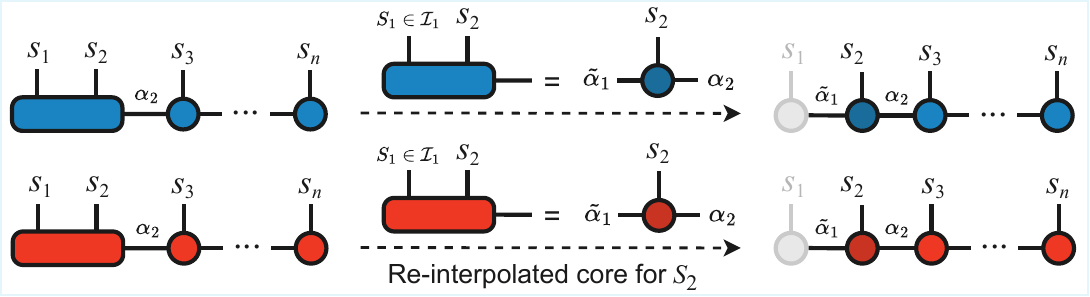}
\end{center}

\subsubsection{Recursion of the Algorithm \label{sec:recursion}}
\label{sec: recursion of the algorithm}

The RSI algorithm proceeds recursively to compute the next TT-core for $G$. In the second iteration, it repeats Steps 1–3, sketching the $s_4$–$s_n$ cores while leaving indices $s_2$ and $s_3$ open; the previously processed index $s_1$ is now excluded. Subsequent iterations follow the same procedure, progressively excluding processed dimensions, opening new ones, and sketching the corresponding segments along the tensor trains. We now discuss several aspects of the subsequent iterations, including slight differences from the initial iteration in the sketching and interpolation steps, as well as the handling of the final iterations and optimizations such as sketch reuse.

A simplified diagram illustrating the subsequent iterations is provided below:
\begin{center}
\includegraphics[width=0.95\columnwidth]{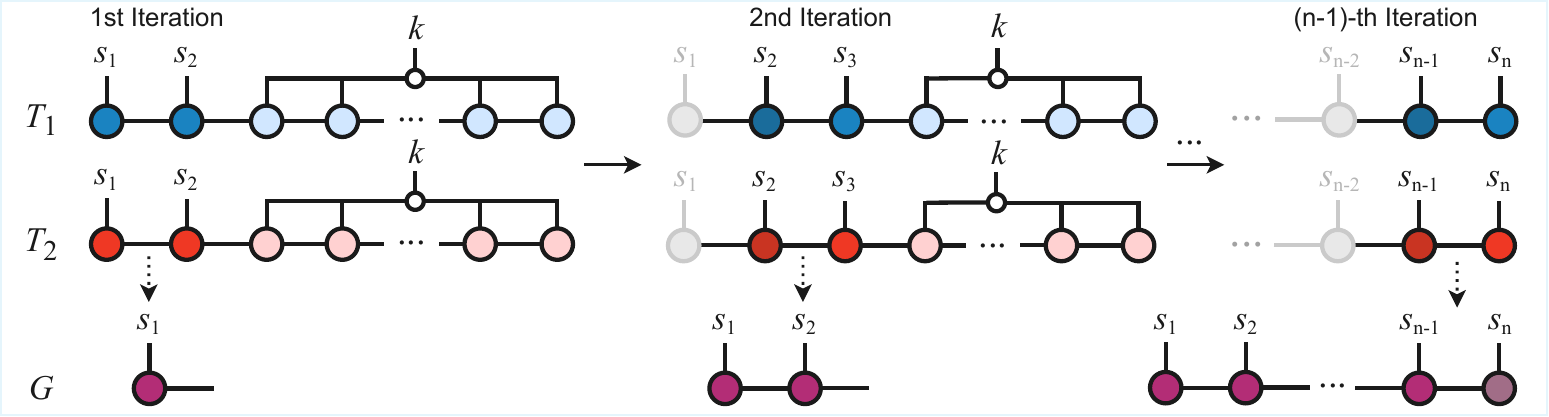}
\end{center}
Since the second iteration, the sketched tensors $\tilde{T}_1$ and $\tilde{T}_2$ have a bond index connecting to the previous TT-core, as well as two external indices and the sketching index. For example, at the second iteration, $\tilde{T}_1$, $\tilde{T}_2$, and their Hadamard product $\tilde{G}$ share the index structure $(\tilde{\alpha}_1, s_2, s_3, k)$, where $\tilde{\alpha}_1$ is the re-interpolated left index with dimension $\chi_1^G$. 
Applying a row ID to $\tilde{G}$ produces an interpolation matrix $X_2$ as the second TT-core and a row pivot set $\mathcal{I}_2$ of size $\chi_2^G$ nested by $\mathcal{I}_1$. The algorithm then re-interpolates the second bonds of the input TTs and continues iteratively in this manner for subsequent cores. 

In total, RSI requires $n-1$ iterations for order-$n$ inputs. Note that the sketching step is required at most up to the $(n-2)$-th iteration, since in the final iteration, as shown at the rightmost of the diagram, only the two TT-cores at $s_{n-1}$ and $s_n$ remain, leaving no further right-side dimension reduction. Hence, the final cores at $s_{n-1}$ and $s_n$ of $G$ are derived by the ID of the Hadamard product of $A^{s_{n-1}}A^{s_n}$ and $B^{s_{n-1}}B^{s_n}$, using its interpolation and skeleton matrix, respectively. Using interpolation matrices for the $s_1$--$s_{n-1}$ cores and the skeleton matrix for the final core, RSI constructs the TT approximation of G in a nested interpolation format.

A key efficiency optimization of RSI involves reusing the KRP sketch to avoid redundant computations. Note that the sketch for the $j$-th iteration is the Khatri‑Rao product of the locally sketched core $\Omega^k_{s_{j+2}}A^{s_{j+2}}$ and the sketch matrix from the $(j\!+\!1)$-th iteration. 
Consequently, after initial generation of all $\{\Omega^k_{s_j}A^{s_j}\}_{j=3}^n$, every sketch matrix for its RSI iteration is obtained by successively applying the KRP between the corresponding sketched TT-core and the sketch matrix for the next iteration:
\begin{center}
\includegraphics[width=0.95\columnwidth]{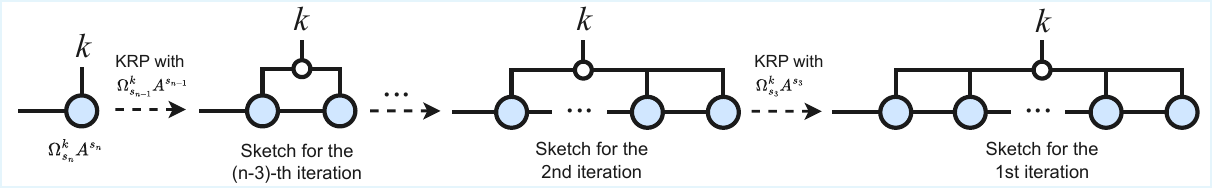}
\end{center}
This enables all sketch matrices to be precomputed in a single backward sweep from $s_n$ to $s_3$ prior to the RSI iterations, stored, and then used directly throughout the iterations.

Another optimization is that sketching can be optionally skipped in the final several iterations. Specifically, at iteration $j$, if the product of the remaining external dimensions, $\prod_{i=j+2}^{n} d_i$, is less than the sketch dimension $k$, then sketching the indices $s_{j+2},\ldots,s_n$ for dimension reduction is unnecessary. In such case, one can omit sketching for the subsequent iterations, and the remaining computation resembles the final iteration's core derivation described earlier: RSI contracts the remaining segment of cores ($s_j$ to $s_n$) from $T_1$ and $T_2$, performs their Hadamard product, and then extracts every TT-core via successive interpolative decomposition over $s_j$ to $s_{n-1}$. The final $s_n$ core is still obtained from the skeleton matrix in the last iteration.

\subsection{Discussion of RSI}
\label{sec: discussion of RSI}

Now we discuss several key aspects of RSI, including the justification and configuration of sketching, its computational complexity, and its relationship to TCI. By analyzing the complexity of a RSI iteration, we show its cubic scaling with respect to the bond dimension.



\subsubsection{Justification of the Sketching Procedure} 


Step~2 applies an ID not to $G$ itself but to the Hadamard product of two independently sketched tensors. We justify here why the resulting pivots still interpolate $G$ accurately. The key point is that the Hadamard product is a diagonal restriction of the \emph{outer} product of the inputs. Define the order-$2n$ tensor
\begin{equation}
  T^{s_1 t_1 \cdots s_n t_n} = T_1^{s_1 \cdots s_n} T_2^{t_1 \cdots t_n},
  \label{eq:outer}
\end{equation}
which, after grouping each index pair $(s_j,t_j)$, is itself a tensor train with cores $A^{s_j}\!\otimes B^{t_j}$. Tying every $(s_j,t_j)$ with the copy tensor recovers $G$. Randomized TT-sketching therefore sketched $T$ directly~\cite{peng2023generativemodelinghierarchicaltensor, camano2025successiverandomizedcompressionrandomized}. Writing $S_1^k$ and $S_2^k$ for the sketches applied to the two inputs in Step~1, and $T'$ for the tensor obtained from $T$ by tying only the open indices $s_1,s_2$, the sketched product of \Cref{eq: sketch G=T1T2} satisfies
\begin{equation}
  \tilde{G}^{s_1 s_2 k} = \sum_{s_3\cdots s_n}\sum_{t_3\cdots t_n}
  T'^{\,\,s_1 s_2 [(s_3\cdots s_n)(t_3\cdots t_n)]}\,
  S^{\,k}_{1\,\,{s_3\cdots s_n}} S^{\,k}_{2\,\,{t_3\cdots t_n}}.
\end{equation}
Taking the Hadamard product after sketching is thus equivalent to sketching $T'$ with the tied sketch $S_1^k S_2^k$, whose site matrices are the one-cluster sketch on the doubled indices $\Omega^k_{1\,s_j}\Omega^k_{2\,t_j}$. Hence $\tilde{G}$ spans the dominant range of $T'$. Since tying $s_j = t_j$ merely selects a subset of the columns of $T'$, the row space of $G$ is contained in that of $T'$, and the sketch supplies a valid basis for interpolating $G$ in the $s_1$ space even though the entries of $\tilde{G}$ differ from those of $G$. The same argument applies to later iterations with the processed indices replaced by the re-interpolated bond index.

This construction addresses the range of $T'$, which can be strictly larger than that of $G$: the row unfolding of $T'$ carries all column pairs $(s_3\cdots s_n)(t_3\cdots t_n)$, whereas $G$ retains only the diagonal subset $s_j = t_j$. A fixed $k$-dimension sketch must therefore spread its resolving power over directions the Hadamard product never accesses, an effect most pronounced for the oscillatory inputs of~\Cref{sec: qtt oscillatory func}, where $T'$ has many comparably weighted directions. It is worth noting that other sketching strategies are conceivable, such as sketching the exact Hadamard product, i.e. the Kronecker product of the TTs, whose feasibility would be worth exploring in future work.

\subsubsection{Sketching Configuration}

By incorporating sketching, RSI introduces the sketch dimension $k$ as a third input parameter, alongside the target maximum bond dimension $\chi_{\max}$ and the ID error tolerance $\epsilon_{\text{ID}}$. Here we assert that determining a good value for $k$ is uncomplicated. In practice, an insufficient sketch dimension can fail to capture enough range of the tensor and degrade the global approximation accuracy. Moreover, it also prevents the output TT from reaching the target bond dimension $\chi_{\max}$.
For example, at the $j$-th iteration, for $\tilde{G}$ with index $(\tilde{\alpha}_{j-1}, s_j, s_{j+1}, k)$, the maximum achievable ID rank is bounded by $\min(\chi_{j-1}^Gd, kd)$. Assuming that $\chi^G_{j-1}=\chi_{\max}$ and $\tilde{G}$ is not rank deficient, we derive a lower bound for the sketch dimension $k$ as $k \ge \chi_{\max}/d$. We therefore set $k$ to exceed this bound by an \emph{oversampling} parameter $p$:
\begin{align}
\label{eq: k vs chimax + oversamp}
k = \frac{\chi_{\max}}{d} + p,
\end{align}
where $p \ge 0$ determines how many extra samples are taken beyond the necessary minimum. Our experiments will show that $p$ can be chosen as a small constant, or even set to zero, with minimal practical impact. 

There are two additional aspects of the sketching configuration worth noting. First, when sketching two TT tensors in practice, the same sketch matrices can be reused for both tensors. Second, the RSI iterations can be designed to leave more than two external indices unsketched. The choice of how many indices to sketch or open reflects a trade-off between accuracy and efficiency. Leaving more external indices unsketched improves the fidelity of the sketched reconstruction to the original tensor but increases its size due to the dimensionality growth. Here our demonstration adopts the most aggressive sketching strategy, which, as will be shown in the experiments, is sufficient to preserve the accuracy of the results.

\subsubsection{Complexity of RSI}
\label{sec: scaling complexity rsi}
We now analyze the computational complexity of RSI. The analysis considers two order-$n$ tensors in TT format, assuming uniform external dimensions ($\{d_j\} = d$) and uniform bond dimensions ($\{\chi_j\} = \chi$) for simplicity. The target bond dimensions for the output are also set uniformly to $\chi$. We also take the sketch dimension as $k=\chi/d$, i.e. no oversampling ($p$=0). This choice is justified because (1) the term involving $p$ is constant and does not affect the scaling with $\chi$, and (2) in practice, $p$ is set to zero or a small value, which the experiments will show is sufficient for RSI. In what follows, we focus on the $j$-th iteration of RSI, whose complexity is representative of all iterations.

\begin{itemize}[leftmargin=*]
    \item \textbf{Step 1 before iteration.} All sketch matrices are precomputed before the iterations begin. This consists of two stages: (1) Applying each random matrix $\Omega$ to its corresponding TT-core, costing $\mathcal{O}(n\chi^2dk)$, which reduces to $\mathcal{O}(n\chi^3)$ when $k=\chi/d$. (2) Successively computing the Khatri–Rao product of the sketch matrix $(\chi\!\times\! k)$ with the preceding sketched TT‑core $(\chi\!\times\! k\!\times\! \chi)$ to form the preceding sketch matrix, proceeding backward from $s_n$ to $s_3$. This requires $\mathcal{O}(kn\chi^2)=\mathcal{O}(n\chi^3/d)$ operations for each input TT. 

    \item \textbf{Step 1 in the $j$-th iteration}. Contracting the sketch matrix with the $s_{j}$ and $s_{j+1}$ cores to form the sketched tensor costs $\mathcal{O}(d^2\chi^3 + d^2 k\chi^2)=\mathcal{O}(d^2\chi^3)$ operations per input TT.
    
    \item \textbf{Step 2 in the $j$-th iteration.} The sketched tensors $\tilde{T}_1$ and $\tilde{T}_2$ have a bond index of size $\chi$, two open external indices $s_j$ and $s_{j+1}$ of dimension $d$, and a sketch index of size $k=\chi$. The Hadamard product of $\tilde{T}_1$ and $\tilde{T}_2$ requires $\mathcal{O}(\chi d^2 k)=\mathcal{O}(d \chi^2)$ operations. Applying a prrLU-based interpolative decomposition to the $(\chi d, \chi )$ matricization of $\tilde{T}_1 \tilde{T}_2$ incurs a third-order cost of $\mathcal{O}(d\chi^3)$.

    \item \textbf{Step 3 in the $j$-th iteration.} Neglecting the cost of slice selection (array indexing), the re‑interpolation step requires contracting two cores for each TT and incurs a complexity of $\mathcal{O}(d^2\chi^3)$. 
    
\end{itemize}

Neglecting the lower cost of the final few iterations (which do not require sketching), all RSI iterations follow the same Steps 1‑3 and share the complexity derived above. Repeating these steps $n-1$ times gives a leading iteration cost of $\mathcal{O}(nd^2\chi^3)$. Together with the sketch precomputation cost of $\mathcal{O}(n\chi^3)$, we can conclude that the overall complexity of RSI is $\mathcal{O}(nd^2\chi^3)$. For TT applications with fixed order $n$ and external dimension $d\ll \chi$, the algorithm achieves leading-order complexity $\mathcal{O}(\chi^3)$, and the bond dimension $\chi$ remains unchanged throughout the entire algorithm.

\subsubsection{RSI and TCI}
\label{sec: RSI and TCI}

Using interpolation—the same principle underlying the TCI method—the RSI algorithm resembles TCI in certain aspects. On a deeper level, it can be interpreted as a variant of TCI in which the deterministic column‑wise slice evaluation is replaced by randomized sketching. Compared with TCI where one treats the input TTs as functions where one queries individual function outputs at $\mathcal{O}(\chi^2)$ cost, the sketched interpolation in RSI avoids the overall $\mathcal{O}(\chi^4)$ cost that would result from using TCI that way. Performed in a single pass, RSI constructs the TT in a left-nested interpolation format through row ID. Additionally, RSI introduces a re-interpolation step, which applies the same pivot selection across input and output tensor trains, thereby unifying their interpolation structure.

The Step 2 of RSI mirrors the underlying logic of TCI, differing in its use of sketched evaluations and single-sided interpolation. One may notice that RSI, like TCI, can be extended to compute more general mappings of TTs. By substituting the product operation in Step 2 with alternative operators, the algorithm can be generalized to broader functional mappings without changing computational complexity. This enables RSI to maintain favorable scalability on more complex tasks, such as computing Hadamard products of multiple TTs or applying element-wise nonlinear functions to a TT, as demonstrated in the experiments in \Cref{sec: Gaussian Functions} and \Cref{sec: nonlinear map of rsi}.

Similar to the cross decomposition in TCI, the ID in RSI terminates when the approximation error falls below the tolerance $\epsilon_{\text{ID}}$ or when the number of selected pivots reaches the target maximum bond dimension $\chi_{\max}$. A larger $\chi_{\max}$ grants the interpolation greater expressiveness, leading to better approximation in both the local ID and the final TT. Note that the ID error is a local measure for $\tilde{G}$ and not a direct measure of the global TT error. These local errors serve as a good proxy for the global error, a common practice in TCI~\cite{Ritter2024QTCI, NunezFernandez2022LearningTrains, Savostyanov2014QuasioptimalityTensors}. As a practical heuristic, setting $\epsilon_\text{ID}$ below the target global TT error provides a useful safety margin for the final result.

\section{Numerical Experiments}
\label{sec: Numerical Experiments}
In this section, we evaluate the effectiveness of RSI by assessing its approximation accuracy and runtime performance in a series of numerical experiments.

\subsection{Experimental Configuration}
We evaluate the approximation accuracy and runtime scalability of RSI across various tensor trains and compare it with conventional methods. The details about baselines and measurements are listed below.
\begin{itemize}[leftmargin=*]
    \item \textbf{Baselines.} We compare RSI against multiple baselines introduced in \Cref{sec: other TT product methods}: 

    (i) The direct method: Forms the explicit Kronecker product of the TTs, followed by TT-rounding to truncate to a target maximum bond dimension or error tolerance.

    (ii) MPO--MPS contraction methods: We adapt the Zip-Up and variational MPO--MPS contraction algorithms to compute the product of two TTs by treating one TT as a diagonal MPO acting on the other. To improve the accuracy of the Zip-Up method, we transform the input MPS and MPO into orthogonal form with the first site as the orthogonality center when running the Zip-Up test.

    (iii) The TCI method: Since only TT-cores are provided, TCI takes as input a function that, given an index, queries tensor entries from input TTs and returns their product.
    Here we use a standard TCI implementation, which sweeps back and forth to find optimal interpolation pivots for each bond and evaluates each queried entry independently, without batching the evaluation of TT environments. The reported TCI runtimes therefore reflect this implementation.
    
    
    \item \textbf{Accuracy.} For small problems where computational resources permit, we quantify the approximation accuracy using the relative Frobenius error between the TT-approximated tensor $\hat{G}$ and the true product $G$, defined as
    \begin{equation}
    \label{eq: rel error}
        \epsilon_r = \frac{\|G^{s_1 s_2\ldots s_n} - \hat{G}^{s_1 s_2\ldots s_n}\|_F}{\| G^{s_1 s_2\ldots s_n} \|_F}.
    \end{equation}
    \Cref{eq: rel error} measures the relative Euclidean distance between $\hat{G}$ and $G$, where a smaller value reflects a more accurate approximation. 
    This global error metric requires contracting TT-cores to evaluate every tensor entry, which becomes infeasible for large problems. For such cases, we instead employ a physically meaningful alternative measurement, as will be detailed in the experiment.
    
    \item \textbf{Runtime.} All experiments were performed on a Linux server (Ubuntu 22.04) with an AMD Ryzen Threadripper PRO 3975WX CPU (32 cores). All algorithms including RSI and other baselines are implemented using the ITensor package ~\cite{itensor} in Julia 1.11.5. Each reported runtime is the average over 5 independent executions.
\end{itemize}

The experiments are conducted on various tensor trains. We begin by computing the diagonal of the density matrix of a quantum many-body system, given a (pure) state wavefunction represented as an MPS (\Cref{sec: product of random TT}).
The demonstration is then extended to quantics tensor trains, which represent continuous functions in a discrete and compressed format, to illustrate the product of functions (\Cref{sec: Function Product in Quantics Tensor Trains}). Due to dependencies on unreleased features within the ITensor library, the Julia implementation of RSI is not yet publicly available. 
But a standalone Python implementation of RSI, along with experiments, is open-sourced at \url{https://github.com/zmeng137/Recursive-Sketched-Interpolation.git}. 

\subsection{Product of Quantum Wavefunction MPSs}
\label{sec: product of random TT}

We address a physics scenario: computing the diagonal of the density matrix associated with a many-body quantum wavefunction represented as a matrix product state. For a pure state $|\psi\rangle = \sum_{s_1,\ldots,s_n} \psi(s_1,\ldots,s_n)\,|s_1\cdots s_n\rangle$, this diagonal corresponds to the joint probability distribution in the computational basis:
\begin{align} 
\label{eq: mps diagonal}
p(s_1,\ldots,s_n) = |\psi(s_1,\ldots,s_n)|^2. 
\end{align}
This distribution encodes measurement statistics necessary for sampling, entropy estimation, and analyzing classical–quantum correspondences. The same squaring operation is the nonlinear bottleneck in solving the Gross–Pitaevskii equation via TT~\cite{boucomas2025quanticstensortrainsolving, SciPostPhys.21.2.035}.
We generate wavefunction MPS using the density matrix renormalization group (DMRG) algorithm: Considering spin‑1 degrees of freedom, we form a matrix product operator (MPO) representing the quantum Hamiltonian
\begin{align}
\label{eq: hamiltonian}
H = \sum_{j=1}^{n-1} \Bigl( S_j^z S_{j+1}^z + \tfrac12 S_j^{+} S_{j+1}^{-} + \tfrac12 S_j^{-} S_{j+1}^{+} \Bigr),
\end{align}
known as the one‑dimensional Heisenberg model. We compute the MPS $\psi$ representing the ground state of the Hamiltonian by applying the \texttt{dmrg} routine from the ITensor library, checking that it is well-converged for each maximum rank or bond dimension used in our experiments. By tuning the \texttt{dmrg} parameters, we generate a set of $\psi$ across a range of problem sizes for algorithm evaluation. 

\subsubsection{20-site System}

\begin{figure}[h]
    \centerline{\includegraphics[width=1\textwidth]{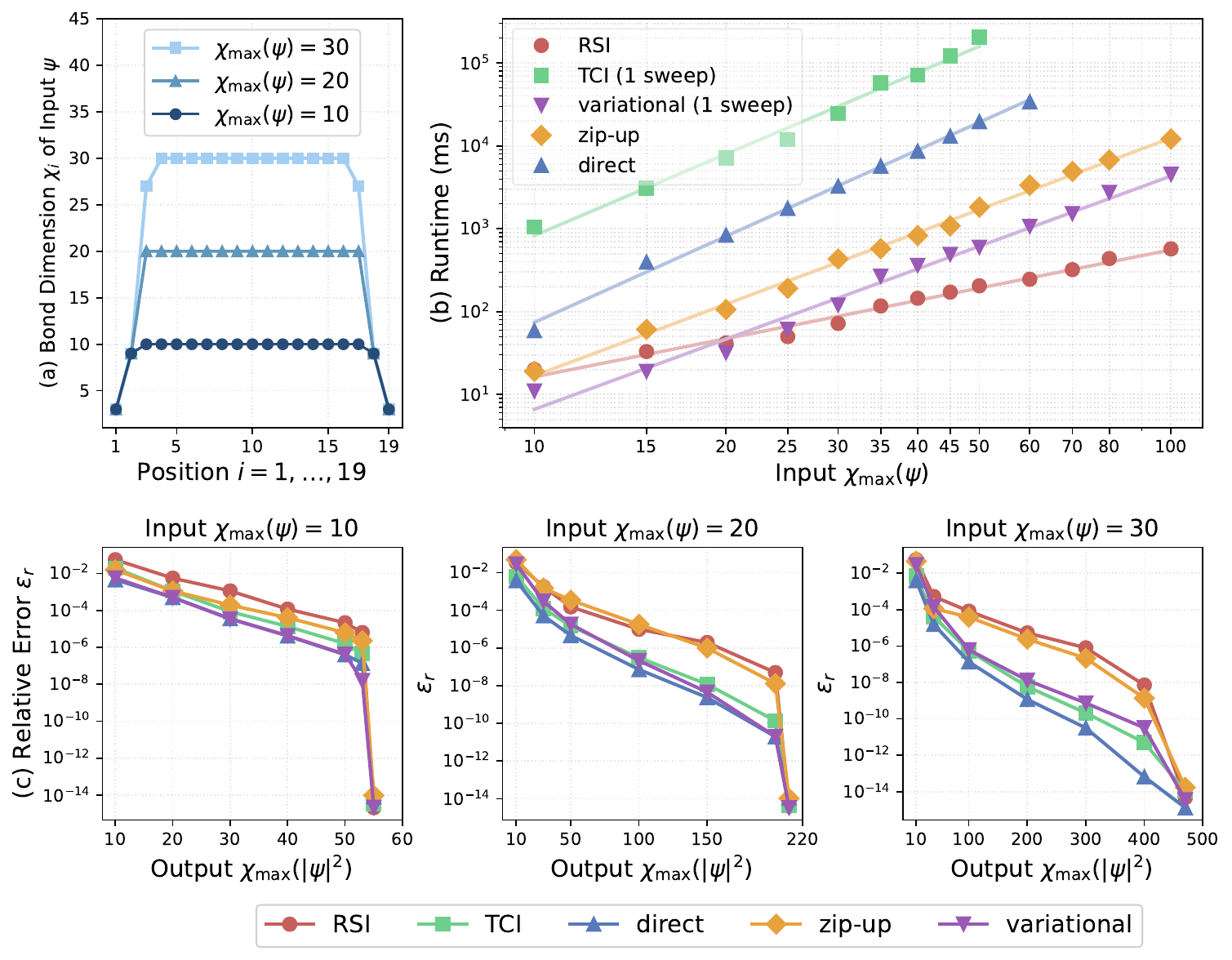}}
    \caption{(a) Bond dimensions $\chi_i$, $i=1,\ldots,19$, of the MPS $\psi$ with $\chi_{\max}=10, 20,$ and $30$. 
    (b) Runtime (ms) versus $\chi_{\max}(\psi)$ for RSI and the baseline methods. The target maximum bond dimension is set to $\chi_{\max}(|\psi|^2)=\chi_{\max}(\psi)$ for all runtime measurements. 
    (c) Relative error $\epsilon_r$ of the product computed by RSI and the baseline methods as a function of the target maximum bond dimension $\chi_{\max}(|\psi|^2)$ for three input cases: $\chi_{\max}(\psi)=10$, $20$, and $30$.}
    \label{fig: n20 mps error runtime}
\end{figure}

We begin with 20‑site systems, i.e. MPS $\psi$ composed of 20 TT‑cores, with various maximum bond dimensions $\chi_{\max}(\psi)$. \Cref{fig: n20 mps error runtime}(a) illustrates the bond dimensions $\chi_i$ ($i=1,\ldots,19$) of $\psi$ with $\chi_{\max}=10$, $20$, and $30$. Owing to the relatively small problem size, we can directly benchmark RSI and other baseline methods on the computation of $|\psi|^2$, evaluating both the relative approximation error $\epsilon_r$ and the runtime for each. 

\Cref{fig: n20 mps error runtime}(c) shows the relative error of $|\psi|^2$ computed by all methods versus the bond dimension of the result, $\chi_{\max}(|\psi|^2)$, for three input cases: $\chi_{\max}(\psi)=10$, $20$, and $30$. To ensure that each method reaches the target $\chi_{\max}(|\psi|^2)$, we set the internal error tolerance to machine precision. For RSI, we use no oversampling, while for TCI and variational methods we perform a sufficient number of sweeps to achieve an optimal approximation. All methods eventually converge to a nearly exact product once $\chi_{\max}(|\psi|^2)$ is sufficiently large, owing to the rank redundancy of the exact Kronecker product of the MPS.

Despite their similar asymptotic convergence, the methods exhibit different accuracy profiles. The direct and variational methods, respecting canonical gauge, consistently achieves the lowest $\epsilon_r$. As the interpolation‑based methods, RSI exhibits a higher error than TCI. This may be because TCI selects  interpolation pivots directly from the true tensor entries evaluated from the inputs over multiple sweeps, while RSI determines pivots from a sketched approximation in a single pass. RSI generally achieves accuracy comparable to, or slightly lower than, Zip-Up with prior canonicalization.

By trading a small amount of accuracy, RSI exhibits significantly more favorable runtime scaling as $\chi_{\max}(\psi)$ increases, as shown in the log--log plot in \Cref{fig: n20 mps error runtime}(b). For runtime measurements, we set the target bond dimension $\chi_{\max}(|\psi|^2)$ equal to the input $\chi_{\max}(\psi)$. For the TCI and variational methods, we report the runtime of a single sweep that fully reaches the target bond dimension. And the runtimes of the direct and TCI methods are reported only up to the largest $\chi$ that could be completed within a reasonable time. Fitting the runtime data to a $\chi^q$ model, we find that RSI exhibits a substantially lower empirical scaling exponent than the baseline methods. With RSI's scaling remaining at least one order lower than the others, the runtime gap between RSI and the baselines continues to widen as $\chi$ increases, providing an initial demonstration of RSI's superior scalability. 


\subsubsection{50-site System}
We now scale the problem to a larger system with 50 sites. Directly computing the approximation error of the 50th‑order tensor becomes computationally prohibitive. Therefore, to assess the accuracy of the approximated $|\psi|^2$, we instead consider physically meaningful scalar observables that can be computed efficiently from tensor networks. In particular, we evaluate the expectation value of the diagonal operator $H_{zz}=\sum_{j=1}^{n-1}S_j^zS_{j+1}^z$, expressed as
\begin{align}
\label{eq: diagonal Hzz}
    \braket{\psi|H_{zz}|\psi} = \sum_{\sigma}|\psi(\sigma)|^2\sum_{j=1}^{n-1}s^z(\sigma_j)s^z(\sigma_{j+1}).
\end{align}
Given the MPO representation of $H_{zz}$ and the MPS of $\psi$, $\braket{\psi|H_{zz}|\psi}$ can be easily calculated using the MPO–MPS contraction and serves as a reference value. 
We evaluate the right hand side of \Cref{eq: diagonal Hzz}, by contracting the $|\psi|^2$ TT with site-wise vectors. The resulting deviation from the reference expectation value is quantified by
\begin{align}
\label{eq: deviation Z}
    Z = |\braket{\psi|H_{zz}|\psi} - \sum_{\sigma}|\psi(\sigma)|^2\sum_{j=1}^{n-1}s^z(\sigma_j)s^z(\sigma_{j+1})|.
\end{align}
A smaller value of $Z$ indicates a more accurate approximation of the probability $|\psi|^2$. 

\begin{figure}[htb]
    \centerline{\includegraphics[width=1\textwidth]{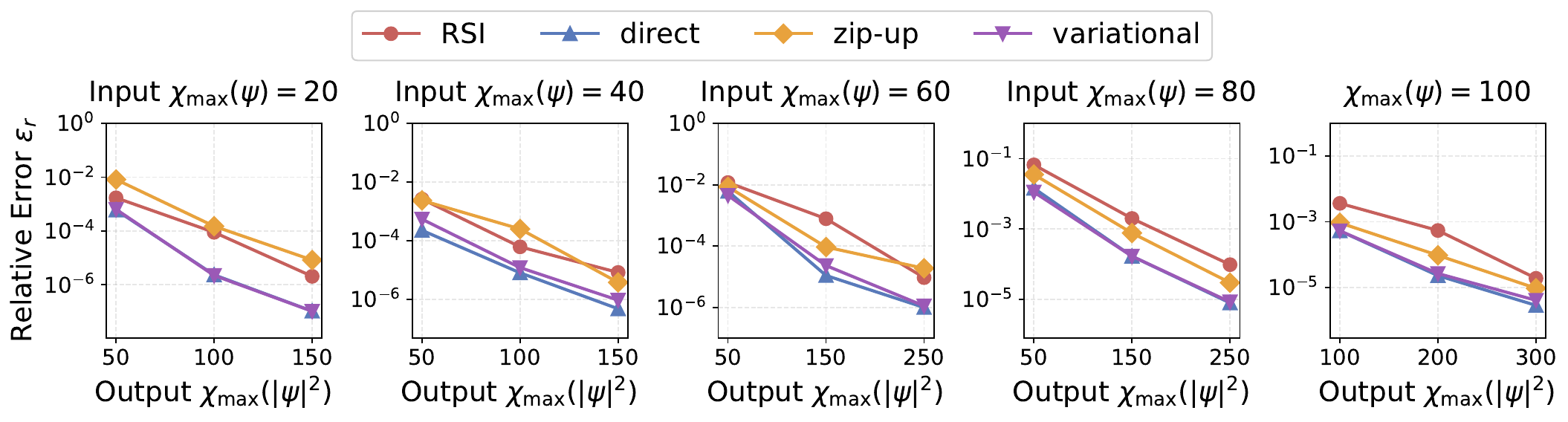}}
    \caption{Convergence of the deviation $Z$ with respect to $\chi_{\max}(|\psi|^2)$ for RSI and other methods, evaluated across wavefunctions with five different bond dimensions.}
    \label{fig: dmrg_mps_expectZ}
\end{figure}

For accuracy measurement via \Cref{eq: deviation Z}, we generate five MPSs $\psi$ with increasing maximum bond dimensions $\chi_{\max}(\psi)=20,40,60,80,100$ using DMRG. For such large systems, our TCI implementation becomes prohibitively expensive to run within a reasonable time. We therefore compare RSI only with the direct method and MPO-MPS contraction methods for computing $|\psi|^2$. From \Cref{fig: dmrg_mps_expectZ}, we observe that, similar to the 20-site system, RSI incurs an average deviation in $Z$ that is approximately one order of magnitude larger than those of the direct and variational methods, while remaining comparable to the Zip-Up method. In return, RSI offers substantially better scalability and significant speedups over the competing methods, making it more practical for computing products of TTs with much larger bond dimensions.

\begin{figure}[htb]
    \centerline{\includegraphics[width=1\textwidth]{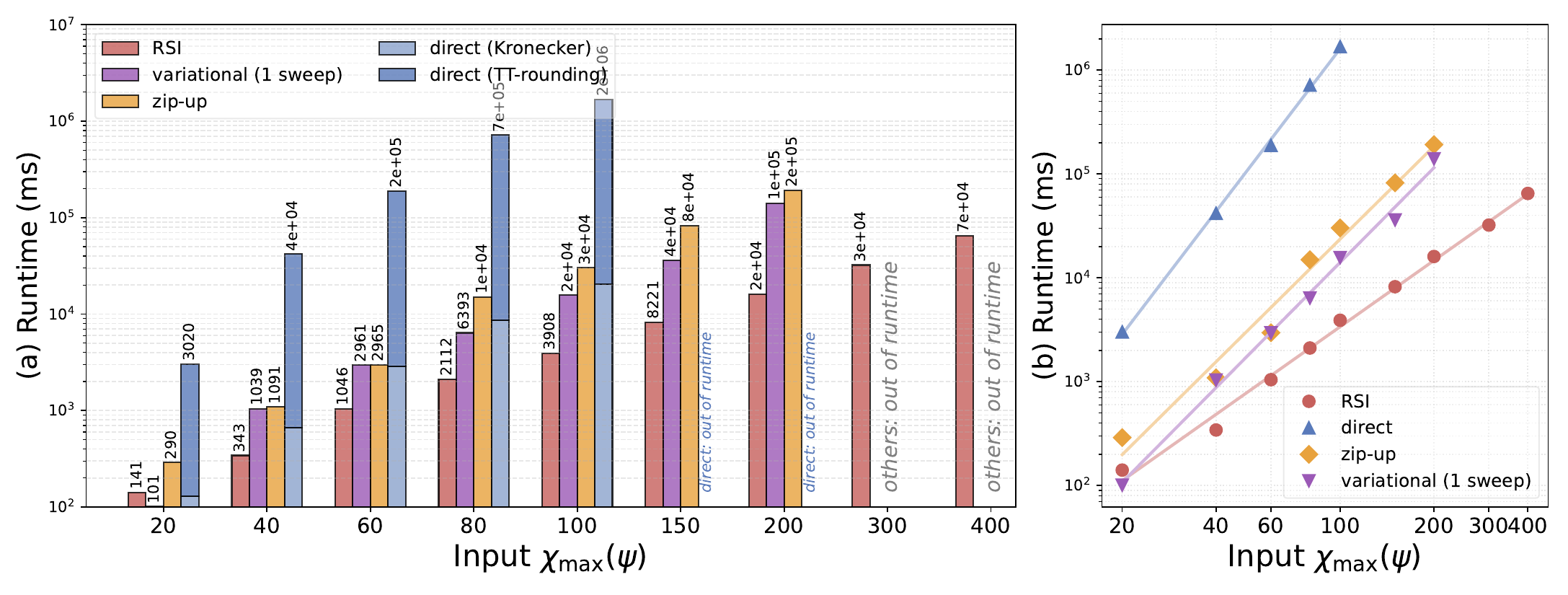}}
    \caption{(a) Runtime comparison of RSI and the baseline methods for input bond dimensions ranging from 20 to 200, with $\chi_{\max}(|\psi|^2)=\chi_{\max}(\psi)$. The direct method is decomposed into Kronecker product computation and subsequent TT-rounding. (b) Empirical runtime scaling of RSI and the baseline methods.}
    \label{fig: dmrg_mps_runtime}
\end{figure}


\Cref{fig: dmrg_mps_runtime} compares the runtime of RSI with the baseline methods through a detailed runtime bar plot in (a) and a scaling comparison in (b). We again set $\chi_{\max}(|\psi|^2)=\chi_{\max}(\psi)$ and measure only a single full-bond sweep of the variational method. In \Cref{fig: dmrg_mps_runtime}(a), we further decompose the runtime of the direct method into explicit Kronecker product computation and TT-rounding, showing that TT-rounding accounts for significantly more runtime than the Kronecker product computation. As $\chi_{\max}(\psi)$ grows, the baseline methods gradually become prohibitively expensive, and we therefore report results only for the largest problem instances that can be completed in a reasonable time.

In comparison, RSI is the fastest method and exhibits the most favorable empirical scaling: fitting the runtime to $\chi^q$ yields $q \approx 2.1$ for RSI, compared with $q \approx 3$ for the MPO–MPS contraction methods and $q \approx 4$ for the direct method. 
The MPO–MPS methods outperform the direct method because they truncate the bond dimension on the fly, whereas the direct method must round an intermediate TT of bond dimension $\chi^2$, which dominates its runtime. 
The fitted exponents of RSI and baseline methods lie below their asymptotic complexities, since over the accessible range of $\chi$ lower-order terms remain significant and optimized linear-algebra kernels of the math libraries achieve higher throughput at larger sizes, so the runtime grows more slowly than the operation count. The difference of roughly one between the exponents of RSI and the MPO–MPS methods is nevertheless consistent with the theoretical gap between $\mathcal{O}(\chi^3)$ and $\mathcal{O}(\chi^4)$. Consequently, the speedup of RSI over the baselines increases with the input $\chi_{\max}(\psi)$, making it the most practical method for large bond dimensions.


\subsection{Product of Functions in Quantics Tensor Trains}
\label{sec: Function Product in Quantics Tensor Trains}

The tensor train can be used to compress the discrete representation of functions. Consider a continuous function $f(x)$ defined on the interval $[0,1)$. To represent this function discretely, one can discretize $x$ via its binary expression:
\begin{equation}
\label{binary x}
\begin{split}
    x^{s_1 s_2 \ldots s_n} & \ =\ 0.s_1 s_2\dots s_n \ \stackrel{\text{def}}{=}\ \frac{s_1}{2} + \frac{s_2}{2^2} + \cdots + \frac{s_n}{2^n}, \\
    s_i &\in \{0,1\} \text{ for } i = 1,\dots,n.
\end{split}
\end{equation}
The discretization in \cref{binary x} yields a uniform grid with $2^n$ grid points and grid spacing $2^{-n}$. The values taken by $f(x)$ on all of these grid points defines the components of an order-$n$ tensor $F^{s_1 s_2 \dots s_n}$:
\begin{equation}
\label{Mode-R F tensor}
    F^{s_1 s_2 \dots s_n} = f(x^{s_1 s_2 \dots s_n}) = f(0.s_1 s_2 \dots s_n) \ .
\end{equation}
The size of this ``tensorized'' representation grows exponentially with the resolution, i.e., the number of binary digits $n$. To achieve compression, one can decompose the tensor $F^{s_1 s_2\ldots s_n}$ into the tensor train as in \Cref{eq: MPS format}, where all physical indices have dimension 2. The resulting TT approximation is referred to as the \emph{quantics tensor train} (QTT) of $f(x)$~\cite{KhoromskijOseledets2010, Oseledets2010Approximation, khoromskij2011d, Jeckelmann}. The compressed QTT representation has wide applications in applications involving high-dimensional functions such as differential equation solving~\cite{boucomas2025quanticstensortrainsolving, dolgov2012FastQttFPEq, Kazeev2014CHEqQTT}. 

\subsubsection{Gaussian Functions}
\label{sec: Gaussian Functions}
We begin with a simple yet common scenario: multiplying Gaussian functions defined as $f(x)= e^{-(x-\mu)^2/2\sigma^2}$, whose structure allows strong compressibility in the QTT format~\cite{lindsey2024multiscaleinterpolativeconstructionquantized}. In the following experiments, we choose quantics digit number $n=25$, discretizing the functions on a $2^{25}$-point grid and yielding QTTs of 25 TT-cores. We obtain nearly exact QTT representations of Gaussian functions ($\epsilon_r \sim 10^{-14}$) using a small maximum bond dimension $\chi_{\max}(f)=10$, and conduct three experiments on such QTTs. We denote the product approximations as $g=f_1f_2$. 


\emph{Experiment 1.} We evaluate the product $g$ of two Gaussian functions $f_1$ and $f_2$ of equal variance $\sigma_1=\sigma_2$, varying the separation between them by adjusting $\mu_1$ and $\mu_2$. For each case, we examine the relative error of the approximated $g$ computed by each method as a function of the target maximum bond dimension $\chi_{\max}(g)$.
\begin{figure}[htb]
    \centerline{\includegraphics[width=1.2\textwidth]{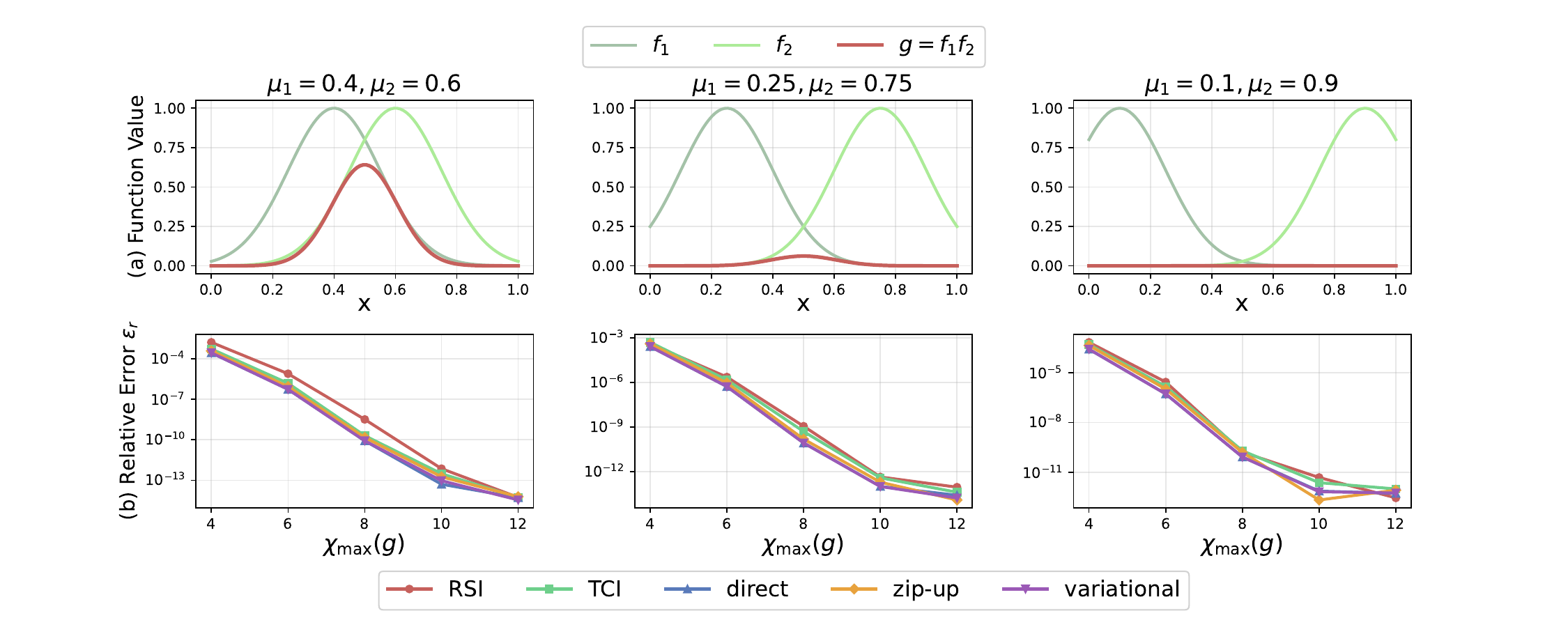}}
    \caption{(a): Shapes of the Gaussian functions $f_1$ and $f_2$ with different $\mu$, and their true product $g=f_1f_2$. (b): Relative error $\epsilon_r$ of the approximated product $g$ derived by different methods versus $\chi_{\max}(g)$.}
    \label{fig: Gaussian function test 1}
\end{figure}
The tests on three different separations in \Cref{fig: Gaussian function test 1} show that RSI achieves comparable error convergence, with relative errors that are either slightly higher than or comparable to the baselines at the same rank. 
As the separation between $f_1$ and $f_2$ increases, $g$ diminishes in magnitude, causing its TT to deviate more significantly from those of the inputs. But this deviation does not impair the ability of RSI to interpolate the structures of $g$ from the TTs of $f_1$ and $f_2$. For each tested pair of means $(\mu_1, \mu_2)$, RSI maintains a similar error convergence trend, achieving $\epsilon_r \sim 10^{-13}$ at $\chi_{\max}(g)=12$. This result confirms that RSI correctly captures the Gaussian structure of $g$ and demonstrates the algorithm's ability to compute products that differ substantially from the inputs.


\begin{figure}[htb]
    \centerline{\includegraphics[width=0.8\textwidth]{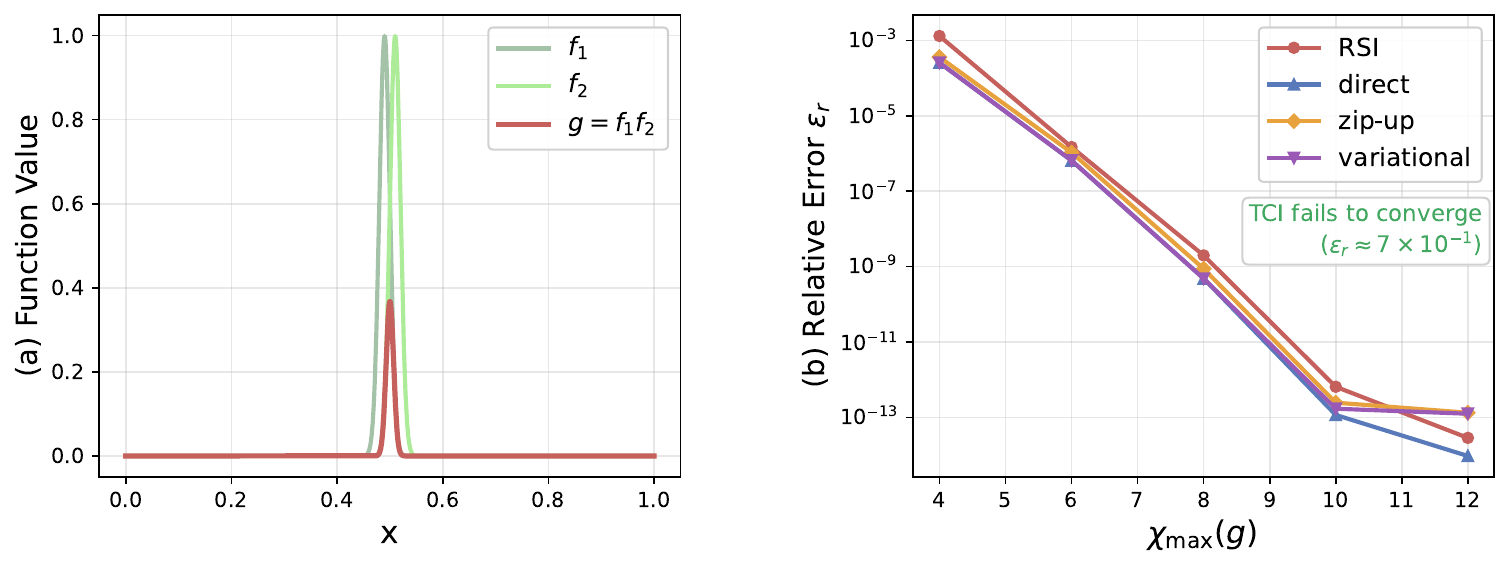}}
    \caption{(a): Shapes of the Gaussian functions $f_1$, $f_2$, and $g=f_1f_2$. (b): Relative error $\epsilon_r$ versus $\chi_{\max}(g)$ of the product approximation.}
    \label{fig: Gaussian spike}
\end{figure}

\emph{Experiment 2.} Beyond the case where the product $g$ becomes increasingly ``flat'' due to a large separation between $f_1$ and $f_2$, another extreme occurs when $g$ becomes highly ``spiked'', which results from inputs with very small variance. \Cref{fig: Gaussian spike} presents the test for this case, computing $g = f_1f_2$ with $\mu_1=0.49,\mu_2=0.51,\sigma_1=\sigma_2=0.01$. The error plot shows that all methods except TCI exhibit convergence behavior similar to that observed in the previous test. We omit the accuracy results for TCI here, as it may fail to converge to an optimal interpolation. Due to the function's sharply peaked structure, which is nearly zero almost everywhere, TCI struggles to locate accurate pivots when starting from a random initial TT, under the same error and rank constraints used for the other two methods. In this test case, we observe that TCI gets stuck at an error level of $\epsilon_r \approx 7\times10^{-1}$ for all values of $\chi_{\max}(g)$.

\begin{figure}[htb]
\centerline{\includegraphics[width=0.8\textwidth]{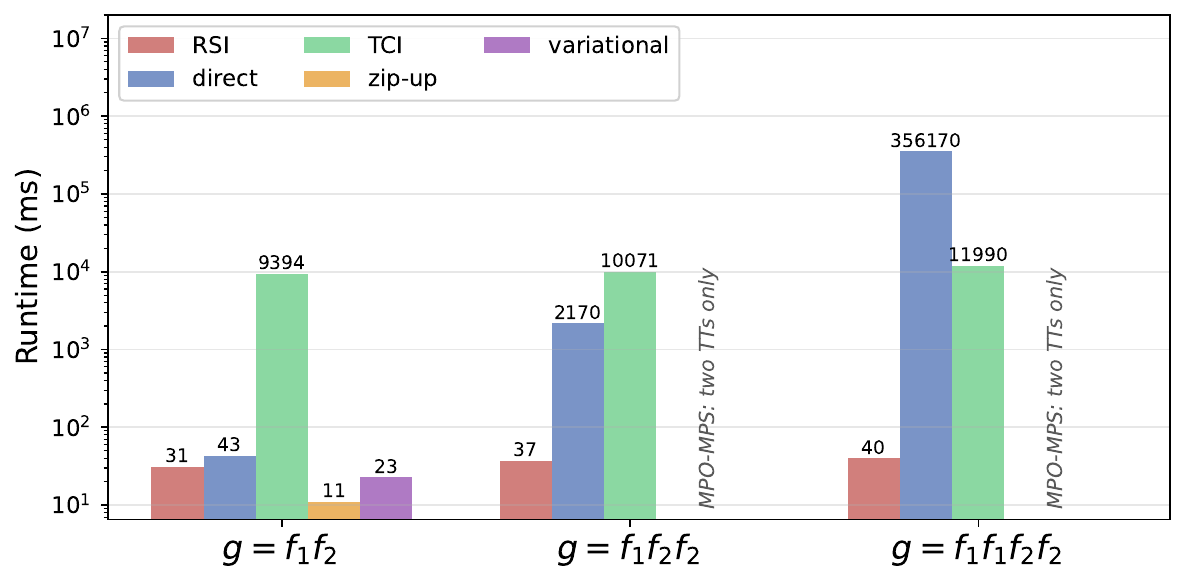}}
    \caption{Average runtime of RSI, TCI, and the direct method in computing $f_1f_2$, $f_1f_2^2$, and $f_1^2f_2^2$.}
    \label{fig: gauss runtime product}
\end{figure}

\emph{Experiment 3.} All preceding experiments can be handled by conventional methods within a reasonable time due to their small bond dimensions and simple product structures. Moving beyond the previous scalability tests with respect to $\chi$, we now use the Gaussian examples to compare the algorithms' scalability for computing Hadamard products involving more than two TTs, such as $g=f_1f_2^2$ and $g=f_1^2f_2^2$. Using $f_1$ with $\mu_1=0.4$, $\sigma_1=0.15$ and $f_2$ with $\mu_2=0.6$, $\sigma_2=0.15$, we apply all methods to compute $f_1f_2$, $f_1f_2^2$, and $f_1^2f_2^2$. The target $\chi_{\max}(g)$ is fixed at $15$, and both TCI and the variational method are performed with two sweeps. Under these settings, all methods achieve sufficiently accurate approximations, with relative errors of $\epsilon_r \sim 10^{-14}$. The runtime of each method is reported in \Cref{fig: gauss runtime product}.

In \Cref{fig: gauss runtime product}, the two MPO--MPS contraction methods are included only for the leftmost product $g=f_1f_2$. For products involving three or more factors, there is no straightforward MPO--MPS formulation. Instead, the product would need to be evaluated through a sequence of pairwise contractions, with truncation applied to each intermediate result. This approach forfeits the benefit of a single global approximation. Therefore, although competitive for the small and simple product case $f_1f_2$, we consider the variational and Zip-Up methods less flexible for computing products involving multiple operands. 

Among the three methods, the direct method scales the worst as the number of multiplied functions grows, due to the exponential increase in cost of both the explicit Kronecker product and the subsequent rounding step. In contrast, TCI treats multiplication as a single function applied to queried tensor entries, making its runtime independent of operations, provided the total number of input TTs remains unchanged. 
For RSI, similar to TCI as discussed in \Cref{sec: RSI and TCI}, adding additional multipliers does not change its $\chi^3$ complexity but only increases the constant prefactor due to the cost of sketching and interpolating extra TTs. In this test case, only $f_1$ and $f_2$ are input, so even the additional cost for extra TTs is absent and only the sketched Hadamard product in Step~2 incurs extra operations. 
Consequently, the runtime of RSI varies only mildly across the three products, making it significantly faster than both baselines. In computing $f_1^2f_2^2$, RSI achieves speedups of $8904\times$ over the direct method and $300\times$ over our TCI implementation.

\subsubsection{Oscillatory Functions}
\label{sec: qtt oscillatory func}

\begin{figure}[htb]
    \centerline{\includegraphics[width=0.9\textwidth]{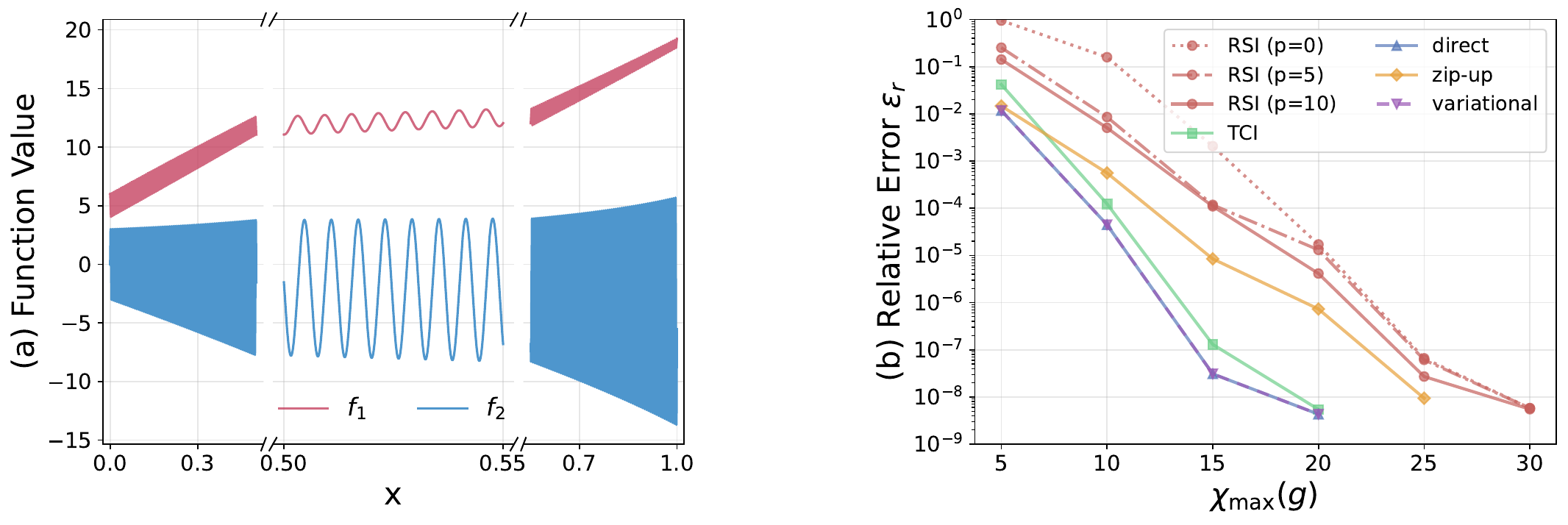}}
    \caption{(a) Oscillatory functions \(f_1\) and \(f_2\). Zoomed-in view on $x\in[0.50,0.55]$ highlights the highly oscillatory structure of the two functions. (b) Relative error \(\epsilon_r\) of the product \(g = f_1 f_2\) approximated by RSI (with various oversampling \(p = 0,5,10\)) and other baseline methods, plotted against \(\chi_{\max}(g)\).}
    \label{fig: osc f1 f2 shape and error cvg}
\end{figure}

The previous experiments demonstrate that, on Gaussian QTTs, RSI achieves significantly better runtime scalability than conventional methods while maintaining comparable accuracy. We now present a case where the input function exhibits a more complex geometric structure that is less amenable to sketching, to examine the algorithm's behavior under more challenging conditions. We consider the multiplication of two highly oscillatory functions, such as
\begin{align}
\label{eq:f1}
    f_1(x) &= \cos\left(2^{10} x \right) \cdot e^{-x^2} + 4e^x - 3x^2 + 10x, \\
\label{eq:f2}
    f_2(x) &= \sin\left(2^{10} x\right) \cdot (e^{x^2}+5x+2) - 4x.
\end{align}
\Cref{fig: osc f1 f2 shape and error cvg}(a) shows the shapes of $f_1$ and $f_2$, which exhibit rapidly varying structures. The quantics representations of such oscillatory functions are known to be well interpolated by TCI into QTT~\cite{Ritter2024QTCI}. Again using $n=25$ quantics digits, we obtain TTs representing $f_1$ and $f_2$ with $\chi_{\max}(f)=10$ and $\epsilon_r\sim 10^{-10}$.

\Cref{fig: osc f1 f2 shape and error cvg}(b) shows the accuracy behavior of RSI and other baseline methods for approximating $g=f_1f_2$ in the QTT format. While the relative error $\epsilon_r$ of all algorithms successfully converges as the target $\chi_{\max}(g)$ increases, RSI converges more slowly and generally requires a larger $\chi_{\max}(g)$ to achieve an $\epsilon_r$ comparable to that of the direct, variational, and TCI methods. Increasing the sketching oversampling $p$ of RSI (from $0$ to $10$) can slightly reduce $\epsilon_r$, but this does not significantly alter the fundamental accuracy gap. With oversampling $p=10$, RSI achieves an $\epsilon_r$ slightly higher than that of the Zip-Up method. 
In particular, when comparing RSI with TCI, we attribute this accuracy gap to the current sketching strategy, since random normal matrix sketching may not be sufficiently expressive to accurately capture the geometry of highly oscillatory tensors, and we sketch the outer product of the input tensors rather than the Hadamard product directly.

When multiplying such oscillatory QTTs, the practical use of RSI requires a slightly increased target bond dimension, which does not affect the overall scalability and still delivers a speedup similar to that shown in \Cref{fig: gauss runtime product}. This example demonstrates that for tensors with more complex structures, the generic random one-cluster basis sketch may capture suboptimal geometry, suggesting a potential for improving RSI through more expressive sketching schemes. We note that there is concurrent ongoing work identifying this limitation of such sketching of TTs in other algorithm scenarios and seeking to enhance its expressibility by introducing structures and increasing the TT-ranks beyond one in sketches~\cite{cazeaux2026linearscalingtensortrainsketching}.

\section{Applications}
\label{sec: applications}

The Hadamard product of TTs arises frequently in real applications. While our numerical experiments benchmarked RSI on quantum wavefunctions, this section briefly demonstrates its broader utility in applications requiring TT-represented function multiplication, such as in nonlinear PDEs and convolutions.


\subsection{Multiplication of Functions in Nonlinear PDEs}
\label{sec: nonlinear pde app}


An important application of TTs is solving differential equations. Recent works leverage quantics tensor trains to compress high-dimensional functions during solving diverse PDEs~\cite{KhoromskijOseledets2010, Kazeev2014CHEqQTT, gourianov2022quantum, peddinti2024quantum, boucomas2025quanticstensortrainsolving}. However, function multiplication in nonlinear PDEs can become a computational bottleneck for QTT-based solvers.
Here we demonstrate the application of RSI for computing QTT function multiplication in PDEs, using \emph{the Well} dataset of PDE solutions~\cite{ohana2024well}. The Well is a curated dataset containing high-resolution solutions to time-dependent PDEs across various domains, providing an easy way to extract specific functions from PDEs for testing. 

\begin{figure}[htb]
    \centerline{\includegraphics[width=1\textwidth]{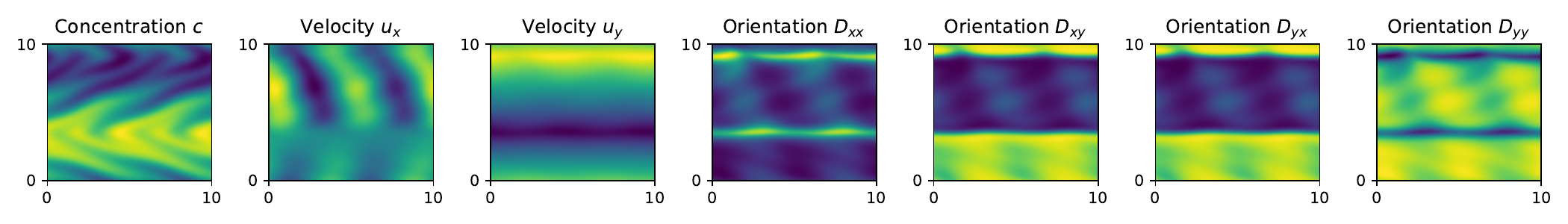}}
    \caption{\raggedright Snapshots of different fields of the active matter system at the 50th time step, including the concentration scalar $c$, the velocity vector $u=(u_x,u_y)$, and the orientation tensor $D=(D_{xx},D_{xy},D_{yx},D_{yy})$.}
    \label{fig: active matter snapshot}
\end{figure}

We select functions from the \emph{active matter} simulation~\cite{maddu2023learningfastaccuratestable}, which implements a continuum theory for the dynamics of rod-like active particles in a Stokes fluid. 
The evolution of the active-matter PDE system involves multiple fields, such as concentration $c$, velocity $u$, and an orientation tensor field $D$. \Cref{fig: active matter snapshot} illustrates snapshots of these fields at the 50th time step on a uniform $256\times 256$ Cartesian grid, from the Well dataset. 
We vectorize these 2D fields into 1D functions on a $2^{16}$-point grid, corresponding to the lexicographic ordering of the Cartesian coordinates, $([x_1,y_1],[x_1,y_2],\ldots,[x_{256},y_{256}])$. The formalized functions exhibit good compressibility under the QTT format, allowing us to construct their QTT approximations with low bond dimensions. 

\begin{figure}[htb]
    \centerline{\includegraphics[width=1\textwidth]{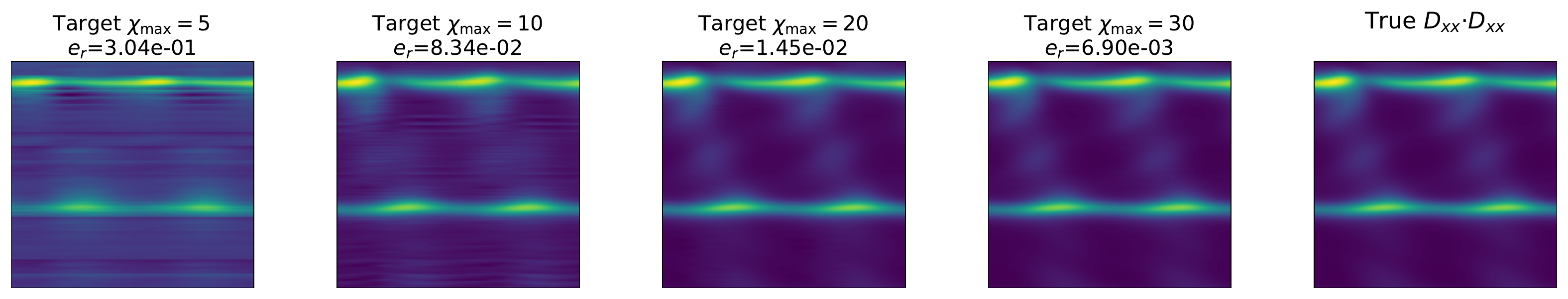}}
    \caption{Approximation of the squared orientation component $D_{xx}\!\cdot\! D_{xx}$ by RSI with various target maximum bond dimensions, alongside the true product. $e_r$ denotes the relative error of the approximate product with respect to the exact product $D_{xx}D_{xx}$.}
    \label{fig: active matter product}
\end{figure}

We demonstrate a representative nonlinear term in the equation, the square of the orientation component $D_{xx}\!\cdot\!D_{xx}$ (see Eq. (7) of \cite{maddu2023learningfastaccuratestable}). We construct the QTT of $D_{xx}$ as input with a maximum bond dimension of 20, achieving a relative error of $2.15\times 10^{-3}$ with respect to the true $D_{xx}$. \Cref{fig: active matter product} illustrates the contours of the true product $D_{xx}\!\cdot\!D_{xx}$ and the approximations produced by RSI for target $\chi_{\max}$ ranging from 5 to 30.  As the bond dimension increases, the RSI result gradually approaches the true product and can reach an accuracy level comparable to that of the input QTT at a rank of 30.



\subsection{Convolution of TT Represented Functions}

Convolution is a fundamental operation that plays a central role in areas such as signal processing and numerical analysis. The convolution theorem states that for two functions, $f_1$ and $f_2$, their convolution can be computed by multiplying their respective Fourier transforms in the frequency domain: $\mathcal{F}(f_1*f_2)=\mathcal{F}(f_1)\mathcal{F}(f_2)$, where $\mathcal{F}$ denotes Fourier transform. For functions represented as quantics tensor trains, efficient algorithms utilizing the matrix product operators have been developed to perform the very efficient Fourier transform on QTTs \cite{Chen_QFT_2023,chen2025direct}. By combining the TT-based Fourier transform with RSI acting as an efficient multiplication operator, we can develop a fast convolution scheme for functions in TT format.

\begin{figure}[htb]
    \centerline{\includegraphics[width=1\textwidth]{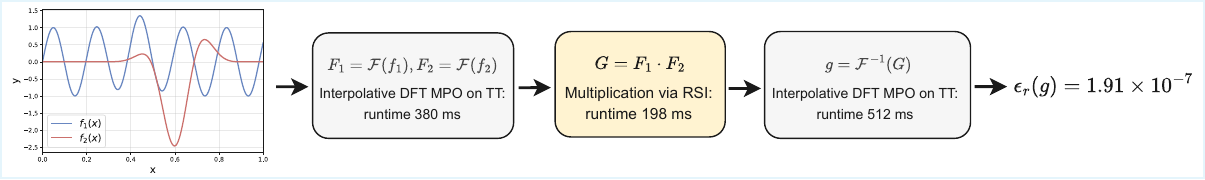}}
    \caption{Workflow of the convolution of two functions in QTT format.}
    \label{fig: conv_workflow}
\end{figure}

Here, we demonstrate a simple case study using RSI for the fast convolution of two QTT functions, the workflow of which is illustrated in Fig.~\ref{fig: conv_workflow}. We initialize nearly exact QTT representations of two functions using 25 cores with a maximum bond dimension $\chi_{\max}=20$. For the forward and inverse discrete Fourier transform operators, we choose to employ an interpolative MPO construction applied to the QTT representations~\cite{chen2025direct}. Using RSI to multiply two TTs representing functions in the frequency domain, we see that the multiplication is not the computational bottleneck in the overall convolution process. The resulting QTT-based convolution approximates the true convolution values with a relative error of $\epsilon_r=1.91\times 10^{-7}$.

\section{Extensions: Nonlinear Mappings of TT via RSI}
\label{sec: nonlinear map of rsi}


In \Cref{sec: RSI and TCI}, we discussed that RSI, similar to TCI, can compute nonlinear operations beyond the Hadamard product of two tensor trains. And as demonstrated in \Cref{sec: Gaussian Functions}, RSI exhibits significant runtime efficiency in computing the Hadamard product involving more than two TTs. In closing, we explore the potential of extending RSI to compute an element-wise nonlinear map $g$ applied to a tensor $T$ in TT format,
\begin{align}
    G^{s_1s_2s_3\cdots s_n}\equiv g(T^{s_1 s_2 s_3 \cdots s_n}).
\end{align}
To accomplish this, RSI follows the same iterative procedure (Steps 1–3) for a given input TT, with the sole modification that computing the Hadamard product in Step 2 is replaced by applying the desired nonlinear function, as illustrated in the diagram below. This modification introduces no additional computational cost compared with the Hadamard product, hence RSI retains its $\mathcal{O}(\chi^3)$ scaling for general nonlinear mappings. 
\begin{center}
\includegraphics[width=0.95\columnwidth]{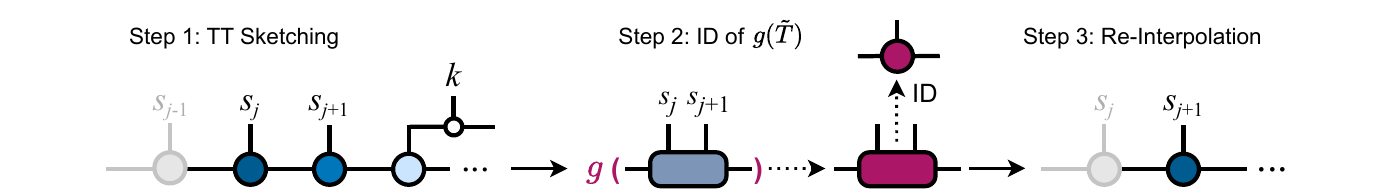}
\end{center}

\begin{figure}[htb]
    \centerline{\includegraphics[width=0.8\textwidth]{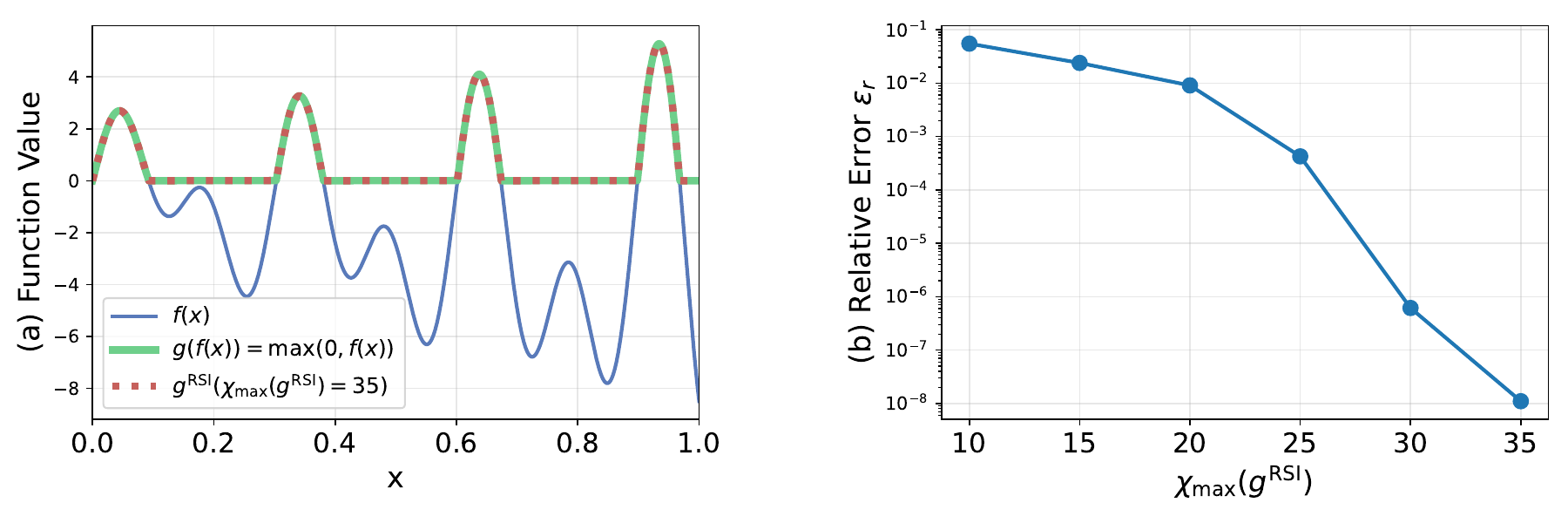}}
    \caption{\raggedright (a) Shapes of \(f(x)\), \(g(f(x))\), and the RSI approximation \(g^{\text{RSI}}\) at \(\chi_{\max}(g^{\text{RSI}})=35\). (b) Relative error \(\epsilon_r\) of $g^{\text{RSI}}$ versus the maximum bond dimension \(\chi_{\max}(g^{\text{RSI}})\).}
    \label{fig: nonlinear map test}
\end{figure}

We provide a brief demonstration of applying a function to a quantics tensor train using RSI. Given a 25‑bit QTT with $\chi_{\max}=10$ that nearly exactly represents a function $f$, we use RSI to apply the famous rectified linear unit (ReLU) function, defined as $g(x)=\max(0,x)$. \Cref{fig: nonlinear map test}(a) shows the shape of $f(x)$, $g(f(x))$; (b) displays the approximation error convergence for $g(f)$ computed by RSI. As the target bond dimension increases, the approximation \(g^{\text{RSI}}\) converges to a low error, accurately reproducing the true function \(g(f)\) at \(\chi_{\max}(g^{\text{RSI}}) = 35\). 

The preceding example, though simple, effectively illustrates the ability of RSI for performing functionals on tensor trains. 
It is worth noting that before RSI, the common method for computing such a mapping of a TT is TCI, which starts from the input TT, applies the mapping to tensor slices, and searches for suitable interpolation pivots. Without knowledge of the true function represented by the input TT, TCI still incurs an $\mathcal{O}(\chi^4)$ cost when tensor slices are queried entry by entry.

Therefore, although a more rigorous mathematical analysis and further experiments lie beyond the scope of this work, we consider that RSI points toward a highly efficient approach for computing nonlinear mappings of tensor networks. 
A detailed study of this extension is left for future work, and it holds considerable potential for a wide range of applications, such as tensorized neural networks or the computation of quantum complexity measures such as coherence or stabilizer Rényi entropy \cite{Kozic_2025, Pan_2025}.

\section{Discussion and Conclusions}
\label{sec: conclusion}

We propose a novel algorithm, Recursive Sketched Interpolation, for computing the Hadamard product of tensors in the tensor-train format with cubic complexity in the TT bond dimension $\chi$. Using randomized TT sketching and interpolative decomposition, RSI attains $\mathcal{O}(\chi^3)$ scaling without forming intermediate large bond dimensions. Numerical experiments demonstrate that by trading off a small amount of accuracy, RSI achieves superior scalability and provides accelerating speedups over conventional methods as the input bond dimension grows.


As discussed, RSI can be interpreted as a variant of the TCI method that substitutes a sketching step for the explicit evaluation of tensor slices, thereby reducing computational complexity. By replacing the Hadamard product in Step 2 with other nonlinear mappings, RSI can serve as a lower-complexity alternative to TCI for applying functionals to TTs. We demonstrate that RSI can handle more complex operations, such as computing the Hadamard product of multiple TTs or applying functions like ReLU to a TT, while maintaining cubic complexity and offering substantial computational advantages.

Several directions for future work naturally follow from the RSI algorithm. 
One concerns the TT sketching discussed in \Cref{sec: qtt oscillatory func}. For TTs with complex geometric structures, the current normal random one‑cluster basis sketch may be insufficient to capture key features, potentially leading to some accuracy loss. Therefore, it would be valuable to explore how alternative sketching strategies could improve accuracy for specific inputs. Another future direction is the generalization of RSI to handle arbitrary element‑wise nonlinear mappings beyond the Hadamard product. Based on the demonstration in \Cref{sec: nonlinear map of rsi} of applying the ReLU function to a TT, we believe it is worthwhile to further investigate this broader use of RSI, as it could enable a wider range of applications for the algorithm.

\section*{Acknowledgements}
We thank Paul Cazeaux, Matthew Fishman, Ivan Oseledets, Marc Ritter, Dmitry Savostyanov, Jan~von~Delft, Nicolas Jolly, Xavier Waintal, and Steven White for helpful discussions.
We thank Qiancan Chen for valuable contributions to the codebase and test cases.
The algorithm was first presented at the 2025 \emph{International Workshop on Tensor Cross Interpolation and Other Algorithms for Learning Tensor Networks}, hosted by the House of Quantum Alps consortium.
EMS would like to thank the organizers and attendees of the workshop for their valuable feedback and discussions.

\bibliography{main.bib}


\end{document}